\documentclass[pra,
article,
twocolumn,
superscriptaddress,
frontmatterverbose, 
amsmath,amssymb,
aps,
floatfix,
10pt
]{revtex4-2}
\usepackage{longtable}
\usepackage[english]{babel}
\usepackage{amssymb,amsthm,amsmath,setspace}
\usepackage{graphicx}
\usepackage{siunitx}
\usepackage[dvips]{epsfig}
\usepackage{geometry}
\usepackage{hyperref}
\usepackage{color,colortbl, multirow}
\usepackage[table]{xcolor}
\usepackage{hhline}
\usepackage{xspace}
\newcommand\moldstruct{M\scalebox{0.8}{OL}DS\scalebox{0.8}{TRUCT}\xspace}
\usepackage{mwe} %Just for example
\addto\extrasenglish{%
}
\usepackage[version=3]{mhchem} % Formula subscripts using \ce{}
\usepackage{subcaption}
\usepackage{caption}

\usepackage{float}
\usepackage{booktabs}
\usepackage{multirow}
\usepackage{siunitx}
\usepackage[percent]{overpic} % in the preamble

\usepackage{xcolor}

\usepackage{comment}

\makeatletter
\def\@email#1#2{%
 \endgroup
 \patchcmd{\titleblock@produce}
  {\frontmatter@RRAPformat}
  {\frontmatter@RRAPformat{\produce@RRAP{*#1\href{mailto:#2}{#2}}}\frontmatter@RRAPformat}
  {}{}
}%
\makeatother

\makeatletter
\expandafter\let\csname longtable*\endcsname\relax
\expandafter\let\csname endlongtable*\endcsname\relax
\makeatother

\begin{document}

\preprint{AIP/123-QED}
\title{%
%\moldstruct: Hybrid Monte Carlo/Molecular Dynamics Simulations of XFEL-Induced Coulomb Explosions%
\moldstruct: benchmarking a hybrid Monte Carlo/Molecular Dynamics model for X-ray free-electron laser ionisation and fragmentation dynamics
}

% Force line breaks with \\

\author{Friederike Krüger}
\affiliation{Department of Physics and Astronomy, Uppsala University, Box 524, SE-75120
Uppsala, Sweden}
\author{Simon Liljeblad}
\affiliation{Laboratory of Molecular Biophysics, Department of Cell and Molecular Biology, Uppsala University, Box 596, SE-75124 Uppsala, Sweden}
\author{Sebastian Cardoch}
\affiliation{Department of Physics and Astronomy, Uppsala University, Box 524, SE-75120
Uppsala, Sweden}
\author{Måns Rosenbaum}
\affiliation{Laboratory of Computational Biology and Bioinformatics, Department of Cell and Molecular Biology, Uppsala University, Box 596, SE-75124 Uppsala, Sweden}
\author{Lasse Eisenach}
%\affiliation{University of Hamburg, Hamburg, Germany}
\affiliation{Department of Physics and Astronomy, Uppsala University, Box 524, SE-75120
Uppsala, Sweden}
\author{Ibrahim Dawod}
\affiliation{Department of Physics and Astronomy, Uppsala University, Box 524, SE-75120
Uppsala, Sweden}
\author{Nicu\c{s}or T\^\i mneanu}
\affiliation{Department of Physics and Astronomy, Uppsala University, Box 524, SE-75120
Uppsala, Sweden}
\author{Carl Caleman$^{*}$}
\email{carl.caleman@physics.uu.se}
\affiliation{Department of Physics and Astronomy, Uppsala University, Box 524, SE-75120
Uppsala, Sweden}
\affiliation{Center for Free-Electron Laser Science,
Deutsches Elektronen-Synchrotron, Notkestraße 85, DE-22607 Hamburg, Germany.}
\author{Tomas André}
\affiliation{Department of Physics and Astronomy, Uppsala University, Box 524, SE-75120
Uppsala, Sweden}

\date{\today}% It is always \today, today,

\begin{abstract}
Single Particle Imaging with intense X-ray free-electron laser pulses requires modelling of the resulting ionisation and Coulomb explosion dynamics of biomolecules to optimise experimental parameters and enable correct structural reconstruction, yet simulating the complete dynamics at protein scale is beyond the reach of quantum-mechanical methods. To address this, we developed \moldstruct, a hybrid Monte Carlo/Molecular Dynamics code built on GROMACS that couples high intense X-ray ionisation dynamics modelled through a Monte Carlo module with classical Molecular Dynamics for atomic propagation.
Benchmarked against quantum mechanical calculations for di-alanine, \moldstruct  agrees in fragmentation patterns above a mean charge per atom of $\bar{z} \approx 1.35$. Compared with Coulomb explosion imaging experimental data for 2-iodopyridine, simulated momentum distributions reproduce the experimental Newton plots in fragment direction and fall within the range of experimental absolute momenta, with a moderate overestimation. We further apply \moldstruct to Protein Explosion Imaging, a method that classifies molecular structures from explosion ion maps recorded on a detector, demonstrating via dimensionality reduction that conformers of the 16-residue peptide $\mathrm{Ala_{16}}$ and five chromophore-labelled ubiquitin mutants are distinguishable. These results establish \moldstruct as a practical tool for simulating radiation damage and Coulomb explosion dynamics of biomolecules at protein scale, where quantum-mechanical methods are computationally infeasible.
\end{abstract}
 
\maketitle
\pagestyle{plain}

Single particle imaging (SPI) is a technique that has the potential to transform structural biology, as it enables the imaging of proteins without the need for crystal growth or freezing~\cite{neutze2000potential}. SPI at X-ray free-electron lasers (XFELs) has been demonstrated for viruses at resolutions below 10~nm~\cite{rose2018single} and, more recently, for the single protein GroEL~\cite{ekeberg2024observation}, yet atomic resolution remains out of reach~\cite{chapman_x-ray_2019}. Several contributing factors have been identified, among them high background contribution~\cite{Daurer2017}, weak scattered signal on the detector and structural heterogeneity of the sample~\cite{mandl}. Significant strides have been made to reduce background contribution~\cite{yenupuri2024helium} and sample heterogeneity~\cite{kierspel2023}, yet achieving diffraction from a single molecule requires high fluencies, like those produced at XFELs, at which rapid multi-photon ionisation drives the molecule into high charge states, causing nuclear dynamics, radiation damage and Coulomb explosion on the fs-timesclae ~\cite{ostlin2018, peptides}. Alongside sample heterogeneity and X-ray shot-to-shot fluctuations, the resulting radiation-induced nuclear dynamics can constrain achievable resolution ~\cite{ostlin2019radiation, gureyev2020radiation, ziaja2012limitations, gorobtsov2015theoretical}. Modelling the complete cascade of XFEL-induced photon- and electron-impact processes, including photoionisation, Auger-Meitner decay, fluorescence, and charge transfer, is therefore necessary to optimise experimental parameters and enable correct structural reconstruction.
% and photon-driven and electron-driven excitations and ionizations

For small systems, this is possible using quantum-mechanical methods such as Density Functional Theory (DFT)~\cite{Aminosyror} or Hartree-Fock-based approaches~\cite{hao2015efficient}, however, such calculations are typically limited to a few hundred atoms, and systems exceeding a thousand atoms remain computationally expensive~\cite{nakata2020conquest}, rendering such approaches infeasible at protein scale.
Continuum plasma models describe XFEL-induced damage through average electron densities and ion charge distributions~\cite{hauriege_dynamics_2004}, but do not resolve individual atomic trajectories or fragmentation patterns. To capture these, hybrid approaches coupling ionisation dynamics with classical molecular dynamics have been developed~\cite{jurek_xmdyn_2016, kozlov_ac4dc_2019, kozlov_hybrid_2020, nass_structural_2020}. \moldstruct follows the same paradigm but is built on GROMACS~\cite{GROMACS}, one of the most widely used open-source molecular dynamics packages~\cite{spoel_gromacs_db_2012, abraham_gromacs_2015}. It directly supports protein data bank structures (PDBs) and provides the preprocessing infrastructure required to set up biomolecular systems, making \moldstruct directly applicable to proteins and other large biomolecules. In previous work, we introduced a collisional-radiative molecular dynamics code for modelling XFEL-induced radiation damage in bulk systems such as protein crystals and liquids~\cite{ionic_liquids, dawod_moldstruct_2024}. Here, we extend the capability of the \moldstruct code to  model single proteins and isolated biomolecules relevant to SPI and single particle structural determination, using a hybrid MC/MD approach.

The model is described in Sec.~\ref{sec:model}. We validate \moldstruct by benchmarking its simulations against quantum-mechanical calculations and experimental data (Sec.~\ref{sec:benchmarking}). In Sec.~\ref{benchmarkmethods} we compare \moldstruct fragmentation patterns and bond-breaking timescales for di-alanine against DFT calculations, in Sec.~\ref{2iodopyridine} we compare simulated with measured momentum distributions, where the experimental data comes from Coulomb explosion imaging of 2-iodopyridine at the European XFEL~\cite{boll_x-ray_2022}. Finally, Sec.~\ref{applications} presents the application of \moldstruct for Protein Explosion Imaging (PXI)~\cite{andre2025,AQUA2018}, a method in which ion fragments from the Coulomb explosion hit a detector and their recorded positions, rather than momenta, encode structural information. Sec.~\ref{sec:ala16} reports that Coulomb explosion ion maps can be used to distinguish conformers of the peptide $\mathrm{Ala_{16}}$, and this approach is extended to chromophore-labelled ubiquitin mutants in Sec.~\ref{sec:ubiquitin}. 
These results demonstrate that \moldstruct provides reliable simulation of large biomolecules such as proteins, currently beyond the reach of quantum-mechanical methods. Accurate modelling of the ionisation-driven dynamics for proteins in SPI experiments enables the systematic optimisation of experimental parameters, such as pulse energy and duration. Furthermore, these findings inform the development of structural reconstruction algorithms that account for radiation-induced nuclear motion.

\section{\moldstruct Model Description}
\label{sec:model}
\begin{figure*}[t] 
\centering
\includegraphics[width=.85\linewidth]{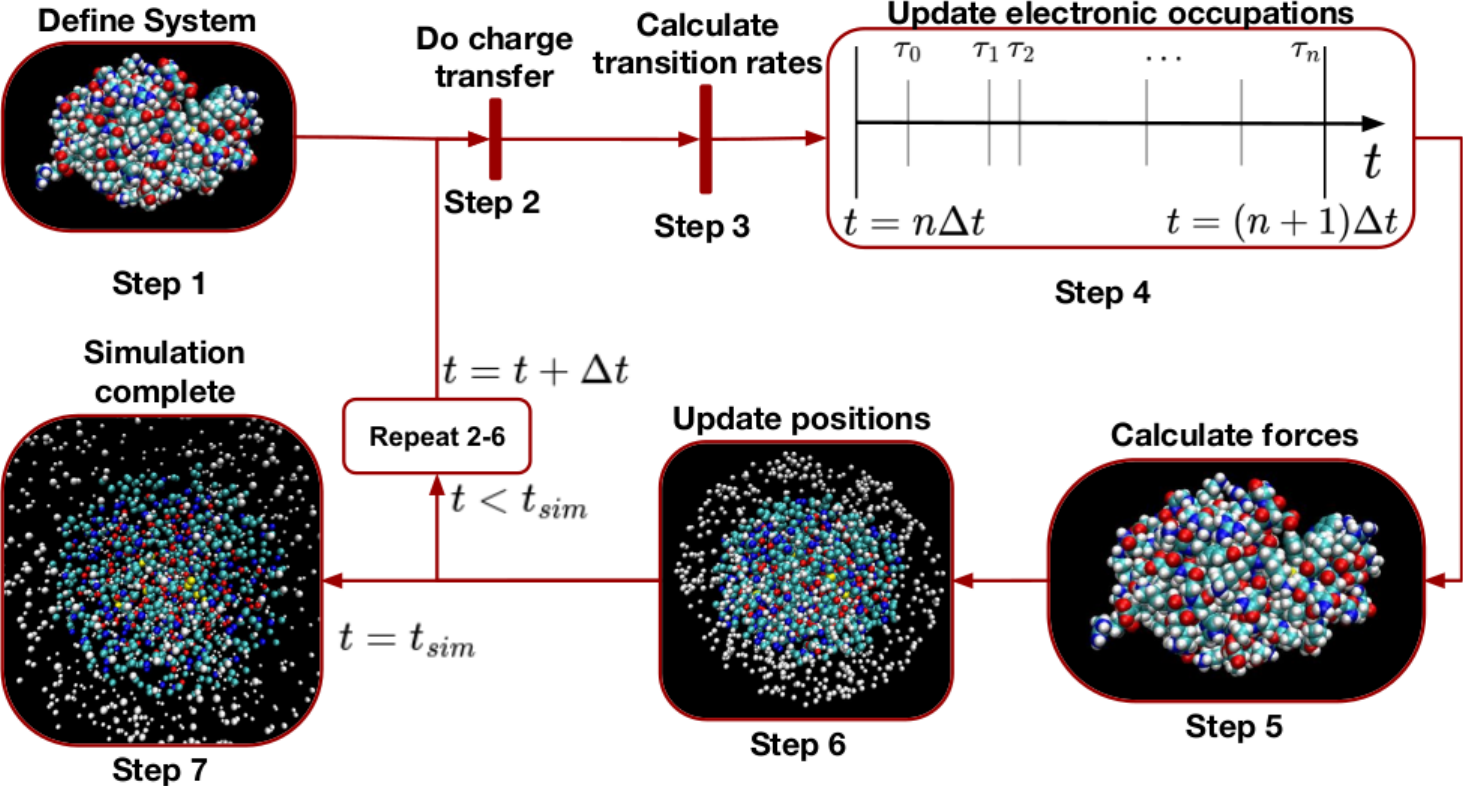}
\caption{Overview of the \moldstruct pipeline, similar to traditional MD but with the inclusion of step 2-4, where Monte Carlo is used to update the electronic occupation of the system.} \label{fig:MC-steps}
\end{figure*}

\moldstruct is a hybrid molecular dynamics/Monte Carlo code for simulations of ultra-fast X-ray dynamics, with an overview of the pipeline shown in ~\autoref{fig:MC-steps}. The code is built on a modified version of the MD software GROMACS~\cite{GROMACS}, augmented with Monte Carlo (MC) algorithms to capture photon-matter interactions. The MC component models the ionisation dynamics and the evolution of electronic states, while the MD component propagates nuclear motion. Between each MD step, a MC module updates the electronic occupation. The code includes photoionisation, radiative and Auger-Meitner decay, and hydrogen charge transfer. We assume photoelectrons and Auger-Meitner electrons escape the molecule before producing significant secondary ionisation. In proteins, the electron inelastic mean free path is approximately 1--10~nm for electron energies in the range 100~eV--10~keV~\cite{tan2006electron}, while the geometric radius of a protein scales approximately as the cube root of its molecular weight~\cite{erickson2009size}. Equating the protein radius with the upper bound of the electron mean free path suggests omitting electron impact holds for proteins up to several hundred kilodaltons in molecular weight. Charge transfer between neighbouring atoms is evaluated via a classical over-barrier model. With updated charges, inter-atomic forces are recomputed and nuclear positions are propagated by the MD integrator.

To compute the charges on each atom \moldstruct uses screened-hydrogenic atomic data~\cite{scott_advances_2010} that is implemented in the collisional radiative code \textsc{CRETIN}~\cite{scott2001cretin}. The atomic data contains level and transition information. Levels are described by their energy, oscillator strength, and statistical weights based on screening constants from More~\cite{more_electronic_1982}. For the transitions, photoionisation cross sections follows Kramers formula~\cite{Kramers_1923_theory}, fluorescence rates use oscillator strength together with statistical weights of the initial and final levels, and autoionisation rates follow the method from Chung~et~al.~(2005)~\cite{chung_flychk_2005}. We include atomic data up to the M shell and compute photoionisation rates every molecular dynamics simulation time step by multiplying the cross section with the X-ray pulse intensity if the photon energy is greater than the ionisation threshold of that level. In principle, the atomic data could also be obtained from other atomic codes or approaches, although this has not been tested.

Within each molecular dynamics time step $\Delta t$, the electronic state of each atom is propagated independently using a kinetic Monte Carlo algorithm. Each transition channel $i \to j$ is assumed to proceed at a constant, independent rate $R_{ij}$, under this assumption, the probability of $k$ transitions occurring in a time interval $t$ follows a Poisson distribution,
 \begin{equation}
     P(R_{ij}, k, t) = \frac{(R_{ij}t)^ke^{-R_{ij}t}}{k!}.
 \end{equation}
Setting $k=0$ yields the survival function $e^{-R_{ij}t}$, the probability that no $i \to j$ transition has occurred by time $t$. Since a Poisson process with rate $R_{ij}$ has exponentially distributed waiting times, the waiting time $\tau_{ij}$ is sampled via the inverse-transform method as $\tau_{ij} = -\ln(U)/R_{ij}$, where $U \sim \mathrm{Uniform}(0,1)$.

The three processes, photoionisation, Auger-Meitner decay, and fluorescence, compete within each MC step: each is assigned an independently sampled waiting time $\tau_{ij}$, and the process with the smallest $\tau_{ij}$ updates the electronic occupations. This is the kinetic Monte Carlo algorithm \cite{gillespie1977, bortz1975}: selecting the smallest $\tau_{ij}$ is equivalent to sampling the next event from the total rate $R_\mathrm{tot} = \sum_{ij} R_{ij}$ and assigning it to process $i \to j$ with probability $R_{ij}/R_\mathrm{tot}$. After each event, waiting times are re-sampled for all processes, including any transition channels that become accessible following the preceding ionisation. This is repeated until the cumulative waiting time exceeds $\Delta t$.

The Auger-Meitner and fluorescence rates are intrinsic atomic properties and do not change within $\Delta t$. The photoionisation rate $\mathrm{PI}_{ij}(t) = \sigma_{ij} I(t)$ depends on the pulse intensity $I(t)$, we therefore hold $I(t)$ constant over each $\Delta t$ and update the photoionisation rates between MD timesteps to reflect the evolving pulse profile.

For modelling the XFEL temporal pulse profile, \moldstruct parametrises a Gaussian pulse by the Full Width at Half Maximum (FWHM) pulse duration, the fluence, and the photon energy.

Hydrogen atoms have a small photoionisation cross section and are rarely ionised directly, instead, they lose their electron via charge transfer to neighbouring atoms. Therefore, charge transfer is implemented for hydrogen donors, describing the transfer of an electron from donor $D$ to acceptor $A$ via a classical over-barrier model \cite{schnorr2014, andre2025}. The potential experienced by the transferring electron is expressed as
 \begin{equation}
    V_{\mathrm{ET}}(\textbf{r}) = - \frac{Q_D+1}{|\textbf{r}-\textbf{r}_D|} - \frac{Q_A}{|\textbf{r}-\textbf{r}_A|},
 \end{equation}
where $Q_A$ and $Q_D$ are the donor and acceptor atoms' charges, $\textbf{r}_A$ and $\textbf{r}_D$ are their respective positions, and $R = |\textbf{r}_A-\textbf{r}_D|$ is the donor--acceptor distance. The energy level of the donor $E_D$ is expressed as
\begin{equation}
    E_D = \epsilon_D - \frac{Q_A}{R},
\end{equation}
with $\epsilon_D$ being the atomic orbital energy of the donor. Along the internuclear axis, the line connecting $D$ and $A$, $V_\mathrm{ET}$ forms a double-well potential with a saddle point between the two nuclei, classically, the electron transfers to the acceptor when $E_D$ exceeds this barrier. Setting $E_D$ equal to the saddle point of $V_\mathrm{ET}$ yields the critical donor--acceptor distance~\cite{schnorr2014, andre2025}
\begin{equation}
    R_\mathrm{crit} = \frac{Q_D+1+2\sqrt{Q_A(Q_D+1)}}{-\epsilon_D},
\end{equation}
and charge transfer occurs instantaneously when $R < R_\mathrm{crit}$.

For the 2-body bonded potential, the standard harmonic potential is replaced by the Morse potential to allow for bond dissociation.  Once the interatomic distance exceeds twice the equilibrium bond length, the Morse contribution is permanently removed and the short-range repulsive term of the Lennard-Jones potential is retained to prevent unphysical overlap of the separated atoms.  An explicit Coulomb repulsion is added between bonded atom pairs to account for the electrostatic repulsion that is otherwise excluded from the non-bonded list.  Additionally, 3- and 4-body bonded potentials are scaled according to the average charge of the involved atoms, ensuring that angle and dihedral interactions gradually vanish upon ionisation.  The scaled potential is given by 
\begin{equation}     
V' = \min(1,\max(1-\bar{z},0))\,V, 
\end{equation} where $V$ is the original potential and $\bar{z}$ denotes the average charge of the atoms included in the potential calculation.  As ionisation removes the bonded terms that the 1–4 Coulomb attenuation compensates for, the attenuation is progressively lifted toward the full, unscaled value. A detailed description of all modifications and a force-field recommendation are given in \autoref{SI:model-description} of the Supplementary Information.

\section{Model benchmarking}
\label{sec:benchmarking}
\subsection{Comparison of \moldstruct with Quantum Mechanics/Density Functional Theory calculations}
\label{benchmarkmethods}
%siesta

\begin{figure*}[t] 
    \centering
    \includegraphics[width=0.8
    \linewidth]{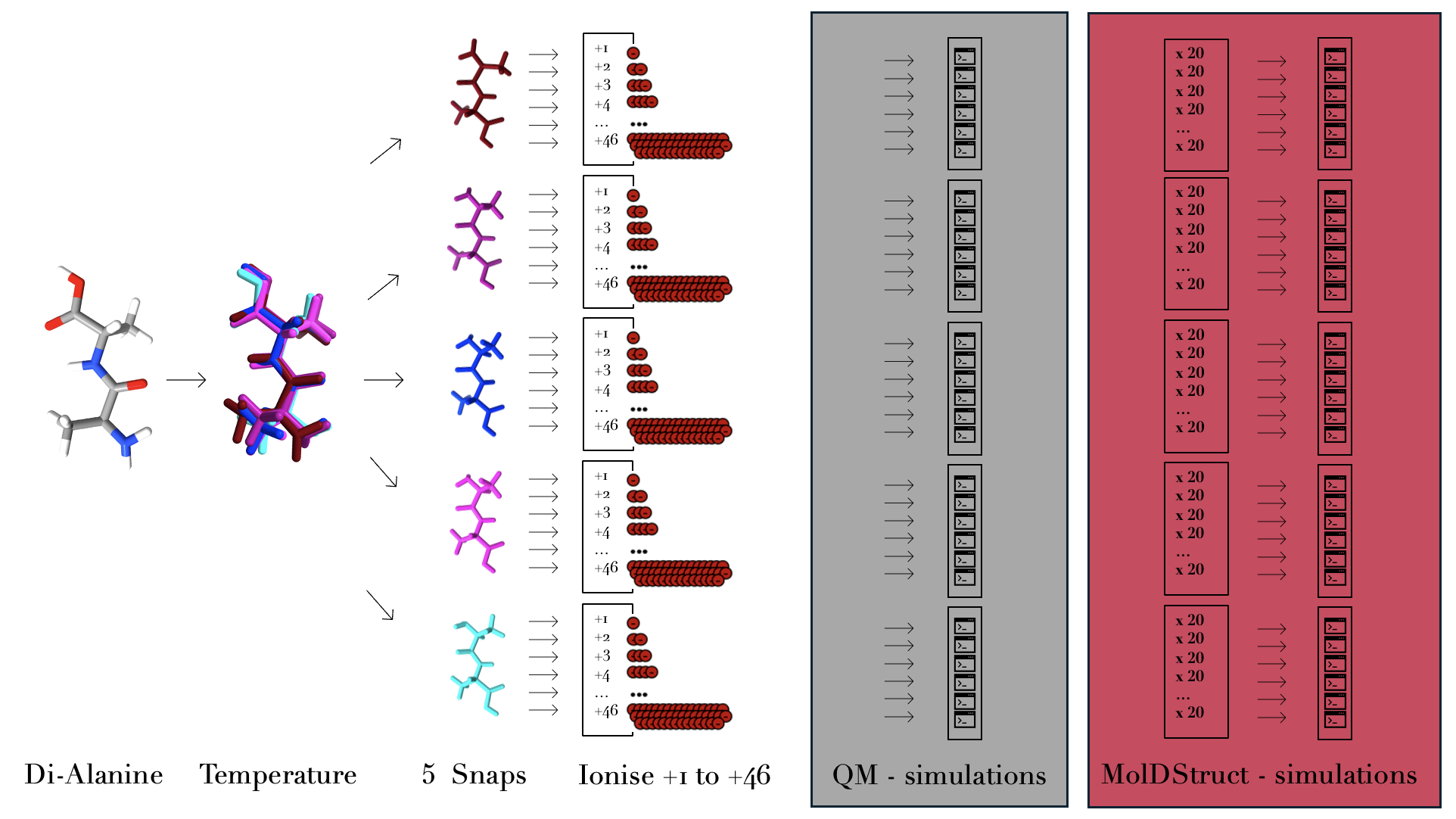}
    \caption{Simulation architecture for the di-alanine benchmark. A short finite-temperature MD run introduces thermal motion into the equilibrium structure, from which 5 independent snapshots are extracted. Each snapshot is then ionised to charge states ranging from $+1$ to $+46$ and propagated independently using both QM (grey) and \moldstruct (red). For each charge state and snapshot, 20 \moldstruct simulations are run to account for the stochastic charge allocation.}
    \label{fig:2Ala_ionisation_scheme}
\end{figure*}

\moldstruct was benchmarked against quantum-mechanical (QM) simulations using SIESTA~\cite{siesta1,siesta2}, which combines Born–Oppenheimer molecular dynamics with DFT. A discussion of the suitability of SIESTA as a benchmark reference is provided in the Supplementary Information~(\autoref{SI:dft-suitability}). Di-alanine was chosen as the benchmark system because QM calculations are computationally demanding, limiting system size (see \autoref{SI:computational-cost} in the Supplementary Information for a quantitative cost comparison), and because alanine is one of the most abundant amino acids in proteins. Bond breaking was recorded as a function of ionisation degree in both methods.
\paragraph{Description of Simulations.}
% It computes the electronic structure of a molecule using DFT, which approximates the solution of the Schrödinger equation for electron density,  and combines it with molecular dynamics to calculated the nuclear motion. It is based on assumptions such as the Born-Oppenheimer approximation and the Kohn-Sham formalism.  
An initial model of di-alanine was constructed using the software Avogadro~\cite{avogadro1,avogadro2} and its geometry was optimized using a conjugate-gradient algorithm within the QM framework (see ~\autoref{fig:2Ala_ionisation_scheme}). 

The optimised structure was subsequently propagated for 500 fs with zero net charge using the generalized gradient approximation and Perdew–Burke–Ernzerhof exchange-correlation functional~\cite{PBE}. This step served to introduce thermal motion into the system. From the trajectory of this propagation, five snapshots of di-alanine were extracted and used as independent starting structures for the fragmentation simulations. Each of these structures was then ionised and propagated using both \moldstruct and QM for charge states ranging from $+1$ to $+46$, with each charge state simulated independently, see ~\autoref{fig:2Ala_ionisation_scheme}.
We defined $\bar{z}$ as the mean charge per atom, where $\bar{z}=2$ is the maximum (full ionisation) for di-alanine (23 atoms, total charge $+46$). Each simulation was propagated for 77~fs (see ~\autoref{fig:2ala_siesta_energy} in the Supplementary Information for the potential-to-kinetic energy conversion).

%siesta simulations
The QM calculations used a time step of 0.5 fs. The electronic temperature and the initial temperature were both set to 300 K for ionisation levels with $\bar{z} < 0.5$. For higher ionisation levels ($\bar{z}>0.5$), the temperature was increased progressively with increasing ionisation, ranging from 600 K up to 8100 K for $\bar{z}=2$. The elevated temperature improved convergence of the electronic structure, while fragmentation patterns remained insensitive to the specific choice of temperature within this range (see ~\autoref{fig:2Ala_temperature_siesta} in the Supplementary Information). This progressive increase in temperature is physically motivated by the larger amount of energy deposited into the molecule at higher ionisation degrees. 
In the applied QM calculations, the total ionisation charge was distributed over the molecular orbitals at the start of the simulation. Partial atomic charges were extracted from the QM simulations to quantify the charge distribution across atoms.

\begin{figure*}[t] 
    \centering
    \includegraphics[width=1.0
    \linewidth]{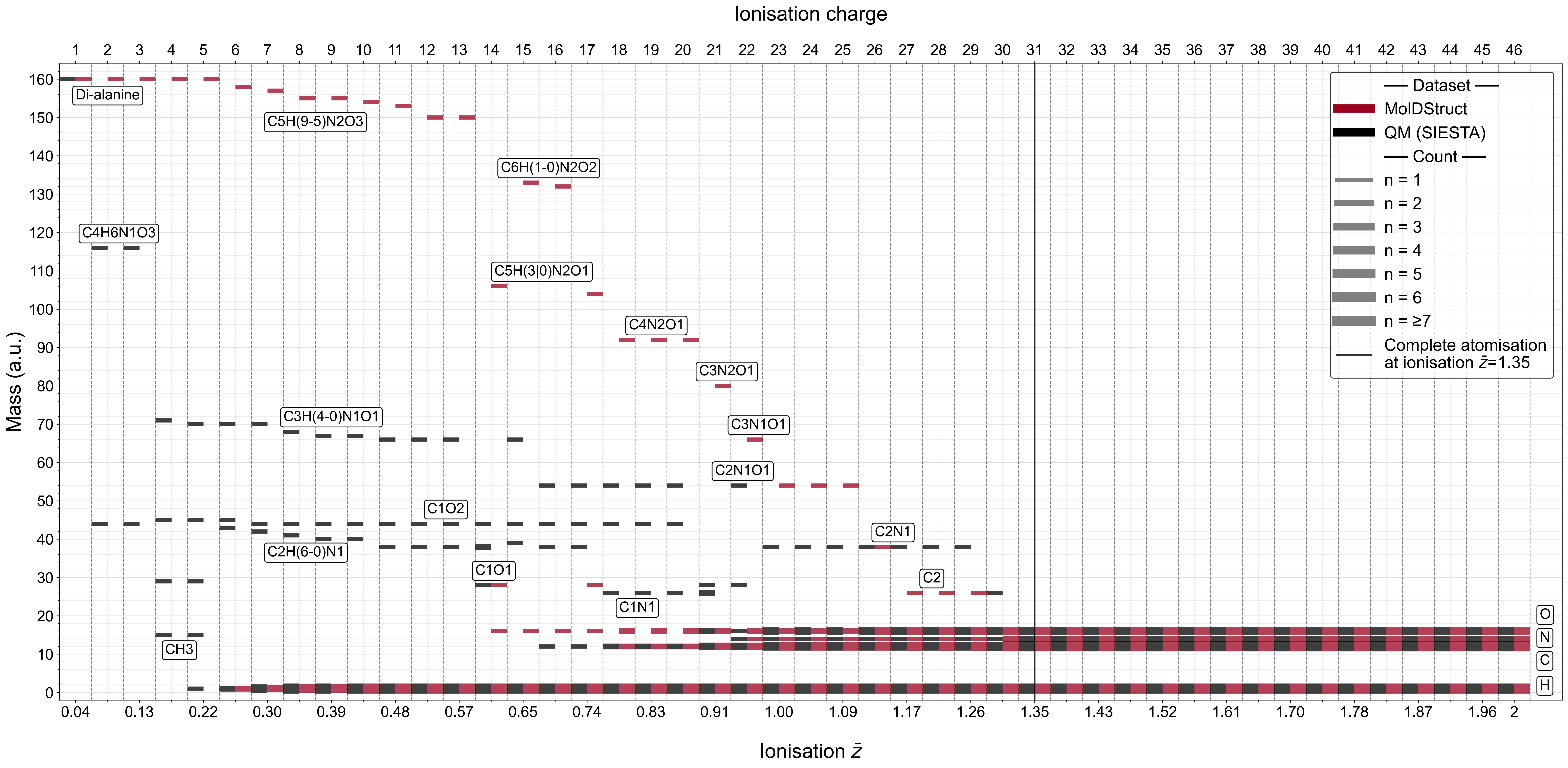}
    \caption{Fragmentation of di-alanine at different levels of ionisation, simulated with QM (black) and \moldstruct (red). The upper x-axis gives the total positive charge (number of electrons removed) and the lower x-axis shows the mean charge per atom $\bar{z}$. The y-axis indicates the fragment mass. The size of each square scales with the number of occurrences of that fragment. Low ionisations (left) leave di-alanine unfragmented, while high ionisations (right) lead to complete atomisation into hydrogen (mass 1), carbon (mass 12), nitrogen (mass 14), and oxygen (mass 16).}
    \label{fig:2Ala_frags}
\end{figure*}

%moldstruct simulations
For this benchmark, \moldstruct charge states were matched to the QM partial charge distribution to enable a direct comparison of nuclear dynamics: integer charges were assigned directly to individual atoms rather than through XFEL-induced ionisation. Auger-Meitner decay, fluorescence, and photo-ionisation were not triggered, though charge transfer remained active to enable the ionisation of hydrogen despite smaller ionisation probablities. Each atom's ionisation probability was set proportional to its QM partial charge relative to the total molecular charge. Charges were assigned sequentially, one electron at a time, by sampling from the QM charge distribution, with each atom excluded once it reached its maximum charge state. The resulting charge configuration was applied instantaneously at the start of each \moldstruct simulation and the system was propagated for 77~fs with a time step of 1~as. Maximum atomic charge states were constrained to the total number of electrons minus two (hydrogen: one ionisation), to avoid unrealistically high charge states in an otherwise post-Auger relaxed system. To account for the stochastic process of ionisation and integer charges, 20 \moldstruct simulations were carried out for every QM simulation and the results were averaged. Bond breaking was assessed using the Souvatzis bond integrity parameter~\cite{bond_integrity}, details of the analysis are provided in \autoref{SI:bond-integrity} of the Supplementary Information.
% \textbf{Discussion}nice

\paragraph{Results and Discussion.}

~\autoref{fig:2Ala_frags} shows the fragmentation patterns of di-alanine as a function of ionisation degree, with \moldstruct results in red and QM results in black. 

Both methods showed distinct fragmentation pathways in the ionisation regime below $\bar{z} = 1.35$. In the regime between $\bar{z} = 1.13$ and $\bar{z} = 1.35$, the results begin to diverge, although matching small molecular species were observed for both methods (see ~\autoref{tab:diala-frags} in the Supplementary Information). Complete atomisation, defined as fragmentation into just single atoms, was reached at $\bar{z} = 1.35$ in the QM simulations and at $\bar{z} = 1.30$ in \moldstruct.

The differences between the results of \moldstruct and QM simulations at the low ionisation regime arise from the fundamental differences in how the two methods treat charge. \moldstruct does not account for molecular orbitals or partial charges on atoms, instead, it assigns integer charges to individual atoms. Furthermore, \moldstruct neglects charge transfers that do not involve hydrogen as the donor. Consequently, \moldstruct may be too simplistic to accurately model low ionisation states, that do not lead to a complete dissociation of the molecule into single atoms but rather rather cause global deformations and minor fragmentation of the molecule.  The gradual ionisation of alanine also resulted in identical fragments from $\bar{z} = 1.31$ onwards for both methods (see Supplementary Information ~\autoref{fig:1Ala_combined}). 
Beyond the fragmentation patterns, we also compared the timescales on which individual bonds broke. The bond integrity of non-hydrogen and hydrogen bonds as a function of both ionisation degree and simulation time is shown in ~\autoref{fig:2Ala_non_H_bond_integrity}  and  ~\autoref{fig:2Ala_H_bond_integrity} of the Supplementary Information. At full ionisation ($\bar{z}=2$), both methods predict near-complete dissociation within $\sim$20\,fs, with closely matching mean bond integrity curves (see \autoref{fig:BI_zbar2} in the Supplementary Information). These heatmaps reveal that at high ionisation levels, where both methods agree on full atomisation, the bond-breaking timescales were broadly consistent between \moldstruct and QM. 

Ion velocity and angular distributions are compared in ~\autoref{SI:velocity-angular} of the Supplementary Information, angular distributions agree closely between \moldstruct and QM for all fragment species, while speed distributions show good agreement for H, C, and N, with \moldstruct overestimating oxygen speeds by approximately 38\%.

\begin{figure*}
    \centering
    \includegraphics[width=\textwidth]{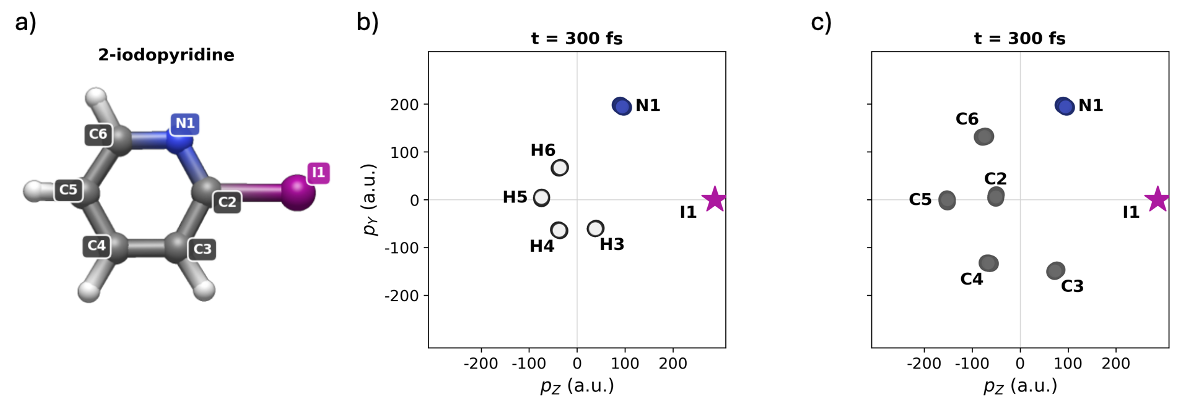}
\caption{\textbf{a)} The molecular structure of 2-iodopyridine. Hydrogen is marked in white, carbon in grey, nitrogen in blue and iodine in violet.
\textbf{b)} Following the Newton-plot convention of Figure 1 b) in \cite{boll_x-ray_2022} this plot shows the simulated momenta of single charged H (white) and  N (blue) fragment ions at time step 300~fs, when the fragment momenta have reached their asymptotic values. Prior to the simulations, molecules were aligned with the molecular ring in the $yz$-plane with the iodine momenta $|\mathbf{p}_\mathrm{I}|$ pointing into positive z-direction, defining the coordinate frame and plotted as a purple star. \textbf{c)} shows the C (grey) N (blue) ions of the same Coulomb explosion, also recorded at $t = 300$~fs.}
\label{fig:2iodpyridine} 
\end{figure*}

\subsection{Comparison with CEI experimental data}
\label{2iodopyridine}
Further, we benchmark \moldstruct against experimental CEI data from Ref.~\cite{boll_x-ray_2022}, collected at the Small Quantum Systems (SQS)~\cite{meyer2022sqs} endstation at the European XFEL~\cite{decking2020mhz}.  In the experimental setup, a femtosecond X-ray laser pulse with a photon energy of 2~keV and an average pulse energy of 1~mJ was focused to a spot size of 1.4~\textmu m. Single 2-iodopyridine, a pyridine ring with an iodine substituent and is composed of 11 atoms ~\autoref{fig:2iodpyridine}a), molecules were injected as a dilute gas into the interaction region and irradiated by the pulse. A cold-target recoil ion momentum spectroscopy (COLTRIMS) reaction microscope (REMI) at the Small Quantum Systems (SQS) beamline was used to measure the three-dimensional momenta of the fragment ions from exploding 2-iodopyridine molecules in coincidence~\cite{boll_x-ray_2022}. An electric field of 423~V/cm was applied at the interaction region, reaching out 86 mm towards the detector. Only singly charged ions (\(q = +1\)) were selectively included in the reconstruction of the experimental Newton plots. 

\paragraph{Description of Simulations.}
With \moldstruct, we modelled the Coulomb explosion of 2-iodopyridine and compared the results to experimental data. Charges of \(+1\) were assigned directly to each atom, resulting in single charged ions, in accordance with the experiment, in which only singly charged ions were selected for experimental reconstruction. In correspondence with the experimental conditions, an electric field of 423~V/cm was applied in $y$-direction, perpendicular to the molecular plane. The ionised system was propagated for 1~ps using a timestep of 1~as. In total, 50 independent simulations were performed with different initial molecular geometries, intended to reflect the thermal motion of the atoms. They were generated from a classical MD trajectory at 5 K temperature starting from the geometry of the original molecule for a time span of 1~ps, sampling a new starting structure every 20 fs. The low temperature reflects the experimental conditions of cold-target recoil ion momentum spectroscopy~\cite{boll_x-ray_2022}.

\paragraph{Results and Discussion.}

The resulting momentum distributions at $t = 300$\,fs,  when the explosion is essentially complete (see ~\autoref{fig:2iod-energy} in the Supplementary Information), are shown in ~\autoref{fig:2iodpyridine} as scatter plots of H (white), C (grey) and N (blue) fragment momenta in the pre-aligned molecular plane, normalised to the iodine recoil momentum $|\mathbf{p}_\mathrm{I}|$ following the Newton-plot convention in Figure~1b (hydrogens and nitrogen) and Figure~1c (carbons and nitrogen) of \cite{boll_x-ray_2022}. Because the simulations were performed in a molecule-fixed frame with the ring pre-aligned in the $yz$-plane, no per-shot frame reconstruction was needed. A full description of the post-processing and normalisation procedure is provided in \autoref{2iod} of the Supplementary Information.
The simulated momentum distributions reproduced the experimental Newton plots of \cite{boll_x-ray_2022} in fragment direction and are in the range of the experimental absolute momenta, with a moderate overestimation. 
This is consistent with two effects: the instantaneous charge assignment in \moldstruct deposits more Coulomb energy than the gradual charge-up during the XFEL pulse, and the average total molecular charge in the experiment was $+9$ for fragmentation into I$^+$ and N$^+$~\cite{boll_x-ray_2022}, reflecting that only singly charged ions are recorded and not all 11 fragments are necessarily detected in coincidence, rather than $+11$ (one positive charge per atom), meaning less Coulomb energy was available for fragmentation than in the simulation. Per-atom momentum components are provided in \autoref{2iod} of the Supplementary Information.
The simulated distributions show a narrower spread than the experimental Newton plots, this may be due to the simulations sampling only 50 initial geometries from a 5~K classical MD trajectory, reflecting a narrow range of thermal configurations, whereas the initial molecular geometries in the experiment are unknown and may span a broader configurational space, resulting in a larger momentum spread. Additionally, the experimental Newton plot reconstruction requires only 4 ions to be detected in coincidence (I$^+$, N$^+$, H$^+$, C$^+$) rather than the full 11-atom complement of 2-iodopyridine~\cite{boll_x-ray_2022}, the charge states of the remaining fragments are therefore unconstrained, introducing additional variability in each explosion and contributing to the observed spread.

\begin{figure*}[htb]
\centering
\includegraphics[width=0.7\textwidth]{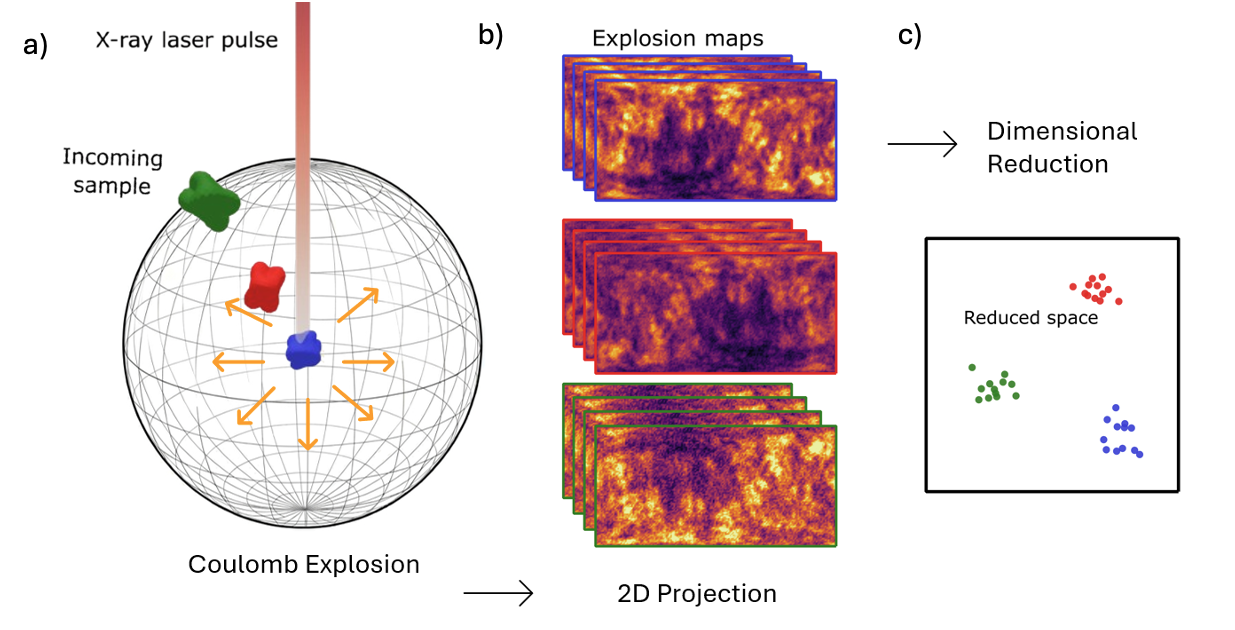}
\caption{Protein Explosion Imaging. \textbf{a)} An ultra-intense X-ray laser pulse hits a molecular sample, triggering multi-ionisation and fragment explosion.
\textbf{b)} Outflying fragment ions are recorded as hits on a 4$\pi$ spherical detector, ion momenta are not directly measured, but the spatial hit distribution encodes the explosion geometry and is projected into 2D spherical ion maps, which act as molecular fingerprints.
\textbf{c)} The ion maps are automatically clustered and visualized in a reduced-dimension space to distinguish different sample states or configurations.}
\label{fig:CEI_scheme}
\end{figure*}

\section{Applications}
\label{applications}

Coulomb explosion imaging (CEI) exploits the Coulomb explosion triggered by an intense XFEL pulse for structural imaging: rapid multi-photon ionisation drives the molecule apart, and the recorded fragment momenta encode the original geometry. CEI typically requires detecting all ionic fragments in coincidence, which limits it to small molecules of a few atoms, however Boll et al. \cite{boll_x-ray_2022} demonstrated that partial coincidences of a fragment subset are sufficient for structural imaging, extending CEI to the 11-atom molecule 2-iodopyridine.

In Protein Explosion Imaging (PXI) ~\cite{andre2025}, instead of resolving fragment momenta in coincidence, only the positions at which ion fragments strike a detector are recorded. Despite this loss of information, the resulting ion-hit footprints remain reproducible for a given molecular structure and thus still encode structural information~\cite{andre2025}.

A schematic overview of a PXI analysis pipeline is shown in ~\autoref{fig:CEI_scheme}. In the following sections, two dimensionality reduction algorithms are applied: \emph{Principal Component Analysis} (PCA)~\cite{PCA}, a linear technique projecting data along directions of greatest variance, and \emph{t-distributed Stochastic Neighbor Embedding} (t-SNE)~\cite{van2008tsne}, a nonlinear technique that preserves local neighbourhood structure in high-dimensional data.

\subsection{Application I: Peptide structure classification}
\label{sec:ala16}

As a stepping stone from the di-alanine benchmark to full protein structures, we apply \moldstruct to $\mathrm{Ala_{16}}$, a 16-residue peptide. The goal is to obtain a range of conformers at varying degrees of structural difference and test whether PXI can distinguish them.

\paragraph{Description of Simulations}
To generate a broad ensemble of conformers, both the $\alpha$-helical and extended conformations of $\mathrm{Ala_{16}}$ were subjected to high-temperature MD simulations at 12,000~K in vacuum using \textsc{GROMACS} with the CHARMM27 force field, in a $4\times4\times4$~nm cubic box with periodic boundary conditions. Applying this extreme temperature induces rapid unfolding, producing continuous trajectories from each starting conformation through progressively more disordered structures over 5~ps (10,000 steps of 0.5~fs). Snapshots were saved every 25~fs, yielding approximately 200 structures per starting conformation spanning a broad range of conformational diversity. A representative subset was selected for PXI analysis, with structural deviation from the $\alpha$-helical reference quantified via RMSD and the local Distance Difference Test (lDDT)~\cite{lddt}, definitions are provided in \autoref{applications_methods} of the Supplementary Information.

\begin{figure*}[htb]
\centering
\includegraphics[width=0.9\textwidth]{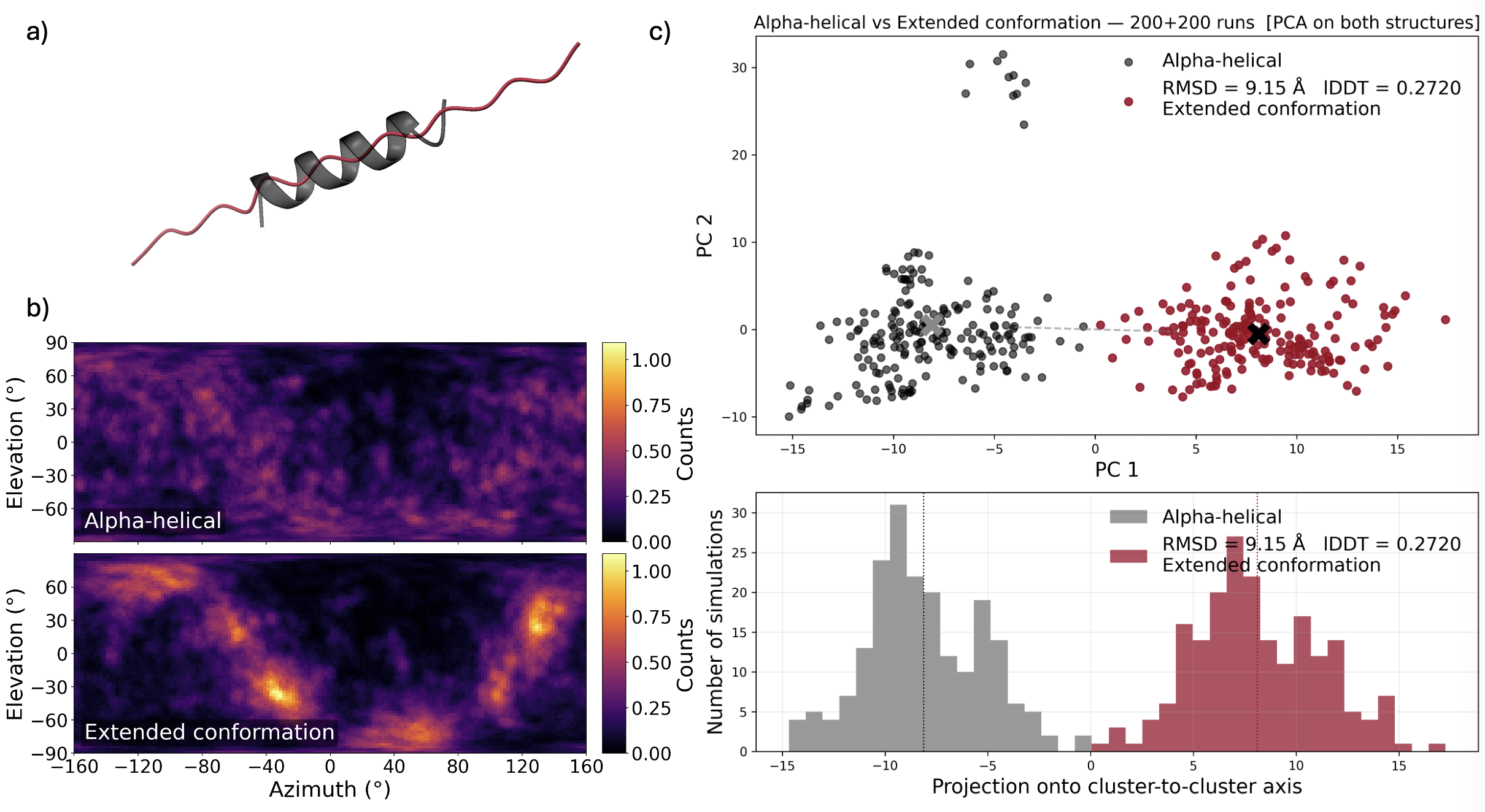}
\caption{\textbf{a)} Molecular structures of the $\alpha$-helical (black) and extended conformation (red) $\mathrm{Ala_{16}}$ peptide conformers.
\textbf{b)} Corresponding Coulomb explosion patterns for the $\alpha$-helical (top) and extended conformation (bottom) structures. Each conformer is subjected to 200 simulated Coulomb explosions at fixed molecular orientation. The emitted ions are recorded on a $4\pi$ detector and the average over all 200 simulations is projected onto 2D ion maps, the colour scale (Counts) indicates the number of ion hits per pixel.
\textbf{c)} Two-dimensional PCA embedding of all 400 Coulomb explosion simulations based on their ion-hit distributions. Each point represents a single simulation, with black points corresponding to the $\alpha$-helical conformer and red points to the extended conformer. Assuming known cluster affiliations, the degree of separation is quantified by projecting the PCA data onto a cluster-to-cluster axis (dashed line). The resulting histogram displays the distribution of these projections, with dotted lines indicating the respective cluster centres to illustrate the clear structural differentiation achieved.
}
\label{fig:16ala}
\end{figure*}

The two conformers used as starting points for unfolding are shown in ~\autoref{fig:16ala}a, the extended conformation has an RMSD of 9.15~\AA\ and an lDDT of 0.2720 relative to the helical conformer, confirming a substantial structural difference. Structures were aligned prior to ionisation.

For the PXI simulations, each selected structure was subjected to 200 independent Coulomb explosion simulations using \moldstruct. The X-ray pulse was modelled with a Gaussian temporal profile with a full width at half maximum (FWHM) of 30~fs ($\sigma = 12.7$~fs), a photon energy of 600~eV, and a fluence of $6.4\times10^{13}$~photons\,\textmu m$^{-2}$. An extended description of pulse parameters is given in \autoref{applications_methods} of the Supplementary Information. The X-ray interaction induced rapid ionisation of the molecule, followed by Coulomb-driven fragmentation, fragment ion hits were recorded on a virtual 3D detector surrounding the interaction region. Dimensionality reduction was applied to the ion-hit distributions.

\begin{figure}[h!]
\centering 
\includegraphics[width=0.47\textwidth]{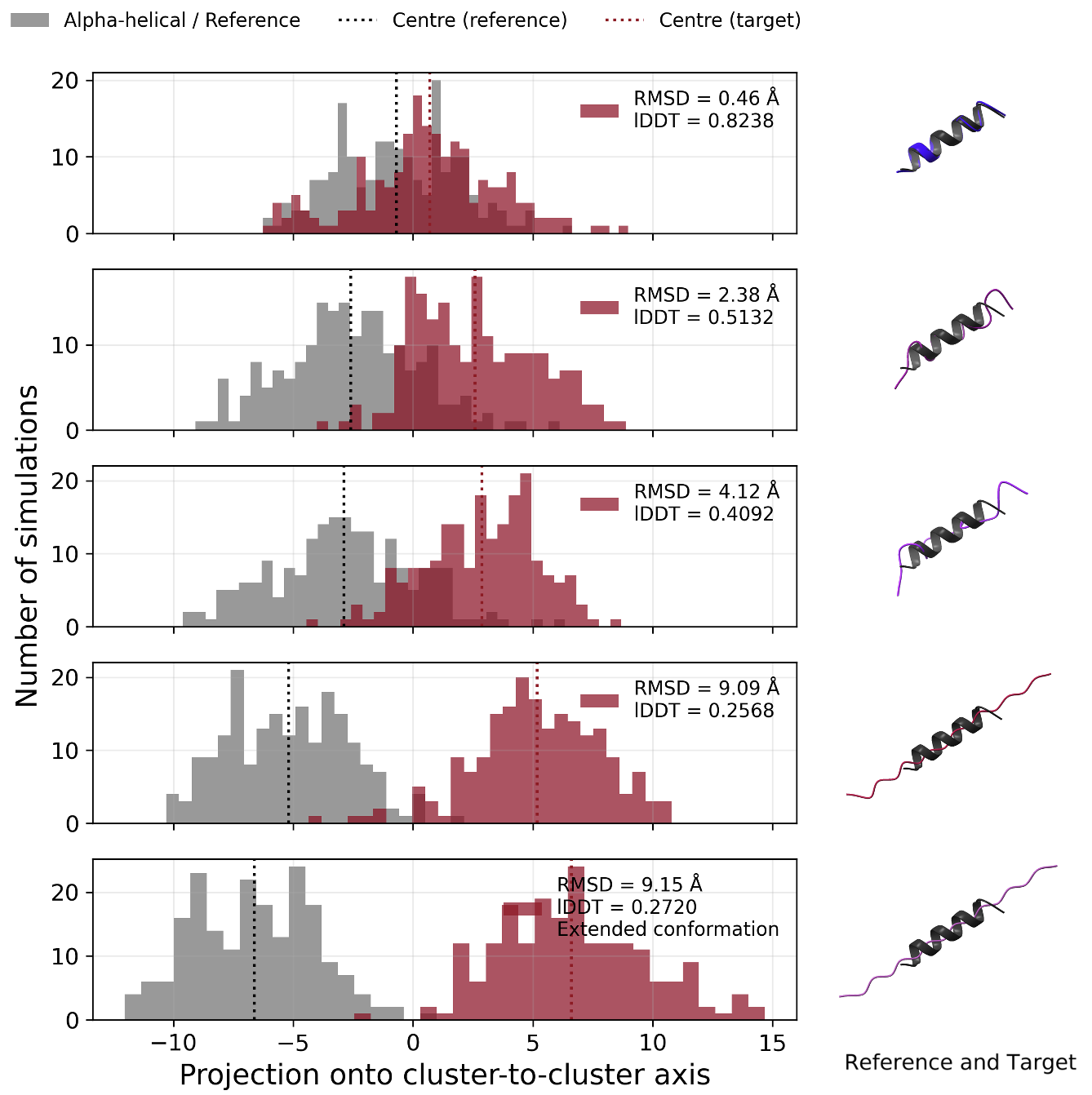}
\caption{On the right, 5 structures of $\mathrm{Ala_{16}}$ are shown together with the $\mathrm{Ala_{16}}$ reference structure. Their structural similarity with respect to the $\alpha$-helical reference decreases from top to bottom, as reflected in a decreasing RMSD value. On the left, the corresponding cluster projections from the PCA analysis are shown, following the same concept as in ~\autoref{fig:16ala}c).}
\label{fig:16ala_tsne_smallsubset} 
\end{figure}

A more extensive set of conformers spanning a broader range of RMSD values is shown in \autoref{fig:tsne_pca_projection_plots_rmsd} of the Supplementary Information.

\paragraph{Results and Discussion.}

\autoref{fig:16ala}a shows the $\alpha$-helical (black) and extended (red) $\mathrm{Ala_{16}}$ conformers. The corresponding Coulomb explosion maps in ~\autoref{fig:16ala}b represent the average of the performed simulations. Both conformers produced the same number of recorded ion hits per simulation, corresponding to 163 atoms of the molecule per run, indicating that the molecule is fully atomised. While the explosion patterns shared some overall similarity, they differed in the spatial distribution of ion hits, encoding the structural difference between the two conformers. Applying PCA to the full set of explosion maps (\autoref{fig:16ala}c) revealed two distinct clusters, one per conformer, showing that the $\alpha$-helical and extended conformers are distinguishable using PXI. Projecting the PCA embedding onto the cluster-to-cluster axis, assuming known cluster affiliations, visualises the gap between the two distributions.

Our results show that $\mathrm{Ala_{16}}$ conformers with an RMSD of 9.15~\AA\ or above from the $\alpha$-helical reference are clearly separable by PXI, with separability increasing monotonically with RMSD, see \autoref{fig:16ala_tsne_smallsubset}, reflecting PXI's sensitivity to global structural differences. For lower RMSD values, that is, $\mathrm{Ala_{16}}$ conformers more similar to the helical reference, separability decreases and becomes less consistent, though increasing the number of simulations per structure may improve discrimination in this regime.
RMSD proves particularly informative here because PXI encodes the global explosion geometry, and a monotonic trend emerges between RMSD and PXI separability — a trend that is less clear when using the local Distance Difference Test (lDDT)~\cite{lddt}, which quantifies local distance differences between atomic neighbourhoods. lDDT values for all conformers are shown in \autoref{fig:tsne_pca_projection_plots_lddt} of the Supplementary Information.
In general, conformers of a small peptide such as $\mathrm{Ala_{16}}$ are harder to distinguish via PXI than those of larger systems, as explored in the following section.

\subsection{Application II: Ubiquitin structure classification}
\label{sec:ubiquitin}
\begin{figure}[htb]
\centering
\begin{minipage}[c]{0.3\columnwidth}
  \centering
  \includegraphics[width=\textwidth]{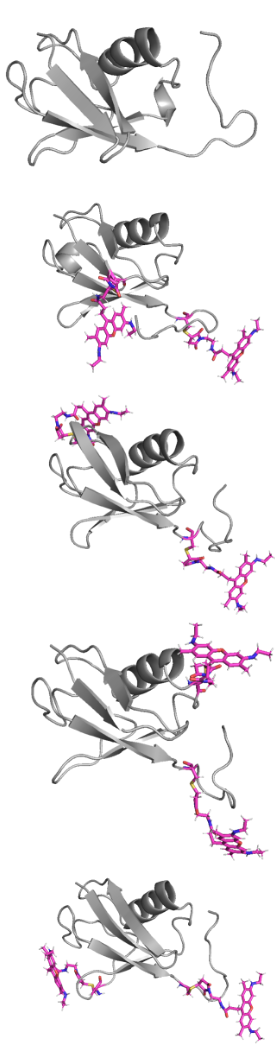}
\end{minipage}%
\hfill
\begin{minipage}[c]{0.69\columnwidth}
  \centering
  \includegraphics[width=\textwidth]{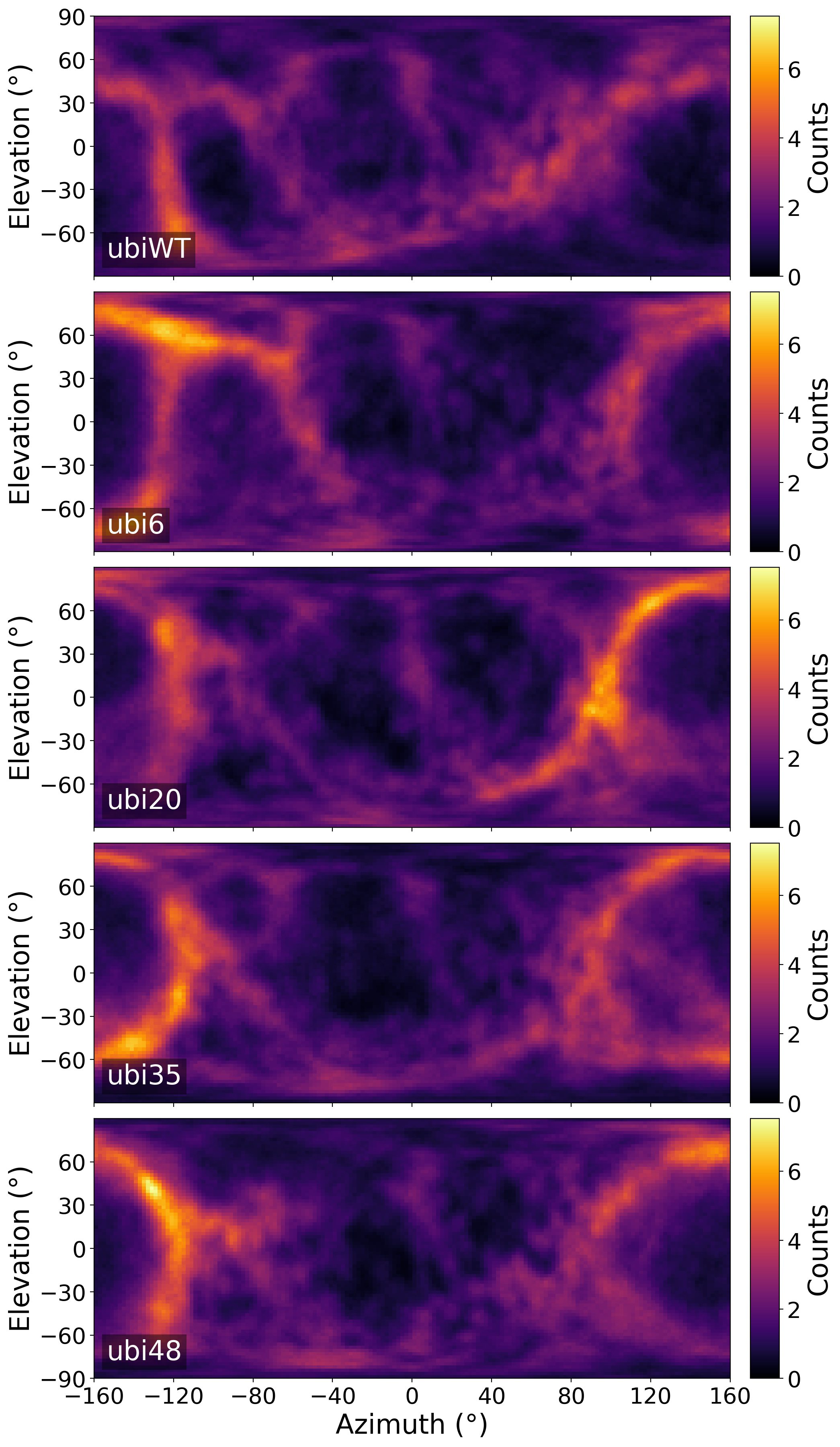}
\end{minipage}
\caption{The ubiquitins used for dimensional reduction. All structures have two chromophores bound at different locations, with the exception of the wild-type. The left side of the image shows their three-dimensional structure, with the protein back-bone coloured in grey and the chromophores in purple. The right side shows each structures respective mean detector image between 100 simulations, with 80x160 bins. Here, the number of total hits varies as different residues are replaced with the ATTO520-cysteine tag, resulting in different numbers of atoms in the molecule and different amounts of hits on the detector.}
\label{ubis}
\end{figure}

\begin{figure*}[htb]
\centering
\includegraphics[width=0.65\textwidth]{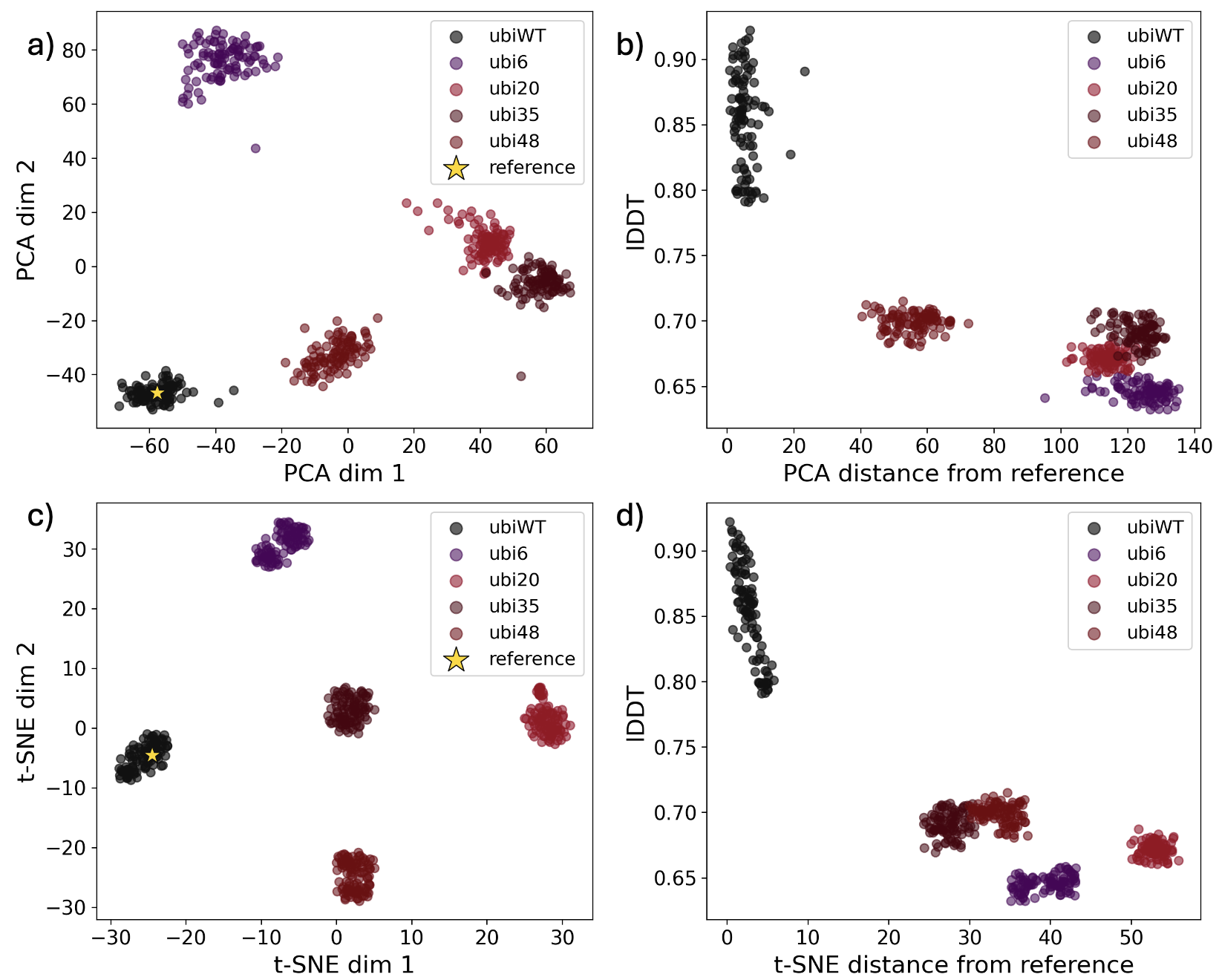}
\caption{Results after performing dimensional reduction. a) A two-dimensional scatter-plot showing all 500 data points after having performed PCA. Every colour corresponds to a specific molecule shown in the legend. The reference corresponds to the centre-most wild-type structure and was used to calculate all lDDT-values and euclidean distances. b) The lDDT-value as a function of the euclidean distance, after performing PCA, to the reference. c) A two-dimensional scatter-plot showing all 500 data points after having performed t-SNE. Every colour corresponds to a specific structure shown in the legend. The reference is the same reference structure used for PCA in subfigure a). d) The lDDT-value as a function of the euclidean distance, after performing t-SNE, to the reference structure. }
\label{PCA_TSNE}
\end{figure*}

Proteins are structurally more complex than small peptides such as $\mathrm{Ala_{16}}$: they are vastly larger and adopt many more conformational states. Here we ask whether PXI can distinguish proteins that differ only in the placement of a single chromophore — a stringent test of structural sensitivity. We use five ubiquitin variants — the wild-type and four chromophore-labelled mutants ubi6, ubi20, ubi35, and ubi48 — shown in ~\autoref{ubis} alongside their respective mean explosion maps. Each mutant carries two ATTO520 chromophores: one always fixed at residue 73, while a second is placed at residue 6, 20, 35, or 48 by replacing that residue with a cysteine functionalised with the chromophore. The chromophores enhance the electric dipole moment of the protein, which is a prerequisite for aligning molecules using an external electric field — a strategy explored to improve the orientation of proteins in SPI 
experiments~\cite{desantis2025ubiquitin}. 
\paragraph{Description of Simulations.}
The explosion patterns for these four mutants, including the wild-type, were calculated for 100  simulations each. All structures were equilibrated in vacuum at times between 200 fs and 50 fs before the pulse hit. This resulted in a total of 500 simulations. The simulated 600 eV pulse had a fluence of $1.27\times10^{13}$~photons\,\textmu m$^{-2}$. An extended description of pulse parameters is given in \autoref{applications_methods} of the Supplementary Information. A spherical detector was used to collect the explosion patterns using $80 \times 160$ bins. Dimensional reduction was applied to all patterns using both PCA and t-SNE in order to yield two-dimensional data points. The euclidean distance between these data points were calculated and plotted alongside the lDDT-values of every structure. These plots can be seen in ~\autoref{PCA_TSNE}.

\paragraph{Results and Discussion.}

The results in figure ~\autoref{PCA_TSNE} show how all five ubiquitin-variants could be separated using both PCA and t-SNE. The distances in the PCA plot ~\autoref{PCA_TSNE}a) were correlated with a lower lDDT-score. The same trend can be seen in ~\autoref{PCA_TSNE}c). Although some patterns overlapped when using a simple dimensional reduction tool like PCA, the more mathematically complex t-SNE managed to separate the ubiquitins.

\section{Discussion and Outlook}

To determine the validity of \moldstruct, its simulations were benchmarked against QM simulations and compared with CEI experimental data from the European XFEL.

Benchmarking of \moldstruct against QM calculations on di-alanine showed that the two methods agree in fragmentation patterns and bond-breaking timescales above a mean charge per atom of $\bar{z} \approx 1.35$, corresponding to the onset of full atomisation. 
Below this threshold, the methods diverged: at low ionisation, partial charges are distributed over molecular orbitals, including charge redistribution through non-hydrogen pathways, which is not captured by \moldstruct's integer charge and atom-localised charge assignment. At high ionisation, the Coulomb forces predominate and the two simulation methods converged. 

A similar atomisation threshold of $\bar{z} \approx 1.31$ was found for alanine, suggesting this is a general feature of the model rather than specific to di-alanine. These results indicate that \moldstruct's approach of simulating through a hybrid MC/MD approach is sufficient to model Coulomb explosion dynamics for ionisations above this threshold.

Taylor et al.\ find that classical simulation methods of Coulomb explosions overestimate ion kinetic energies and underestimate the spread of angular distributions compared to time-dependent DFT~\cite{taylor2025}, that study concerns IR laser-driven Coulomb explosions with TDDFT as the quantum reference, both differing from the XFEL context and ground-state DFT reference used here. Ion velocity and angular distributions, compared with ground-state DFT in ~\autoref{SI:velocity-angular} of the Supplementary Information, show that \moldstruct agrees closely with DFT for angular distributions of all elements and for ion speeds of C, N, and H, although the hydrogen speed distribution is broader in the simulations; for oxygen, a systematic peak offset is observed, as discussed therein. The experimental comparison reinforces these findings: simulated Newton plots for 2-iodopyridine (Figure~\ref{fig:2iodpyridine}) reproduce the directional structure of ion fragments and are in the range of the experimental absolute momenta, with a moderate overestimation. Because only singly charged ions were selected in the experiment and unit charges were assigned to all atoms in the simulations, both operated at $\bar{z} = 1$, this agreement therefore validates \moldstruct in the $\bar{z} = 1$ ionisation regime. Together, these results bracket the reliable operating range of \moldstruct: the QM benchmarking confirms full reliability above $\bar{z} \approx 1.35$, while the experimental agreement at $\bar{z} = 1$ demonstrates applicability even at moderate ionisation.

We presented two applications demonstrating the potential of combining \moldstruct with dimensionality reduction for structure classification via PXI. For $\mathrm{Ala_{16}}$, conformers with an RMSD of 9.15~\AA\ or above from the $\alpha$-helical reference were clearly separable by PCA of their explosion patterns, with separability increasing monotonically with RMSD. For ubiquitin, all five variants, differing only in the placement of a chromophore, were fully separated by t-SNE, with PCA providing partial separation. The correlation between Euclidean distance in the reduced space and RMSD further confirmed that the explosion maps encode genuine structural information. Together, these results establish \moldstruct as a practical computational tool for predicting the structural discriminating power of PXI experiments prior to their execution. Combining SPI and PXI in experiments could be advantageous, as additional structural information from PXI, such as orientation recovery \cite{AndrePartial2025,andre2026orientation}, could support data processing and help overcome the signal-to-noise limitations that currently prevent atomic-resolution SPI. Furthermore, MolDStruct simulations could help to understand the electronic and nuclear damage during an XFEL pulse. By knowing the ionisation state as well as the nuclear displacement of each atom, one could compute time-resolved structure form factors in the process of refining the electron density in SPI experiments~\cite{yoon2016comprehensive}.

Several approximations in \moldstruct limit its accuracy in specific regimes. The independent-atom approximation neglects molecular orbital effects and partial charges, charge transfer is modelled only for hydrogen donors, and secondary ionisation by free electrons is not included. These limitations are least consequential in the high-ionisation Coulomb explosion regime for which the code was designed, and where the benchmarking confirms its reliability.

Future work could extend charge transfer beyond hydrogen donors, incorporate more realistic XFEL pulse parameters directly into the ionisation scheme, and apply \moldstruct to larger systems such as viruses or nucleic acids.

\section{Conclusion}
\moldstruct is a hybrid MC/MD code built on GROMACS for simulating the interaction of high-intensity XFEL pulses with molecules, and it is available on GitHub\footnote{\href{https://github.com/moldstruct/mc-md}{github.com/moldstruct/mc-md}}. Benchmarking against QM calculations established reliable operation above a mean-per-atom ionisation of $\bar{z} \approx 1.35$, while experimental validation against CEI data for 2-iodopyridine demonstrated good agreement even at $\bar{z} = 1$. \moldstruct has been used to show that explosions carry unique footprints corresponding to the molecular structure~\cite{andre2025}, demonstrating the possibility of PXI as well as the opportunity to use it for hit detection during SPI experiments. Additionally, \moldstruct can be used trace explosions and use them to infer the orientation of samples during delivery~\cite{AndrePartial2025,andre2026orientation}, which could serve as a critical step in SPI experiments  for scarce or weakly scattering samples. By providing reliable predictions of radiation-induced dynamics on nano-scale samples, \moldstruct offers a practical route to optimise experimental parameters and inform reconstruction strategies ahead of experiments, supporting the broader effort toward atomic-resolution single particle imaging of proteins and nanoparticles.

\section{Acknowledgments}
Project grants from the Swedish Research Council (2018-00740, 2019-03935, 2023-03900) are acknowledged, and the Helmholtz Association through the Center for Free-Electron Laser Science at DESY. 
CC acknowledges support from a Röntgen Ångström Cluster grant provided by the Swedish Research Council and the Bundesministerium für Bildung und Forschung (2021-05988).  This work was supported by the European Union SPIDOCs Marie Skłodowska Curie Action program, HORIZON-MSCA-2022-DN Grant Agreement Nr. 101120312.

The computations were enabled by resources in projects NAISS 2024/5-140, 2025/5-375, 2024/22-768, 2025/22-862  provided by the National Academic Infrastructure for Supercomputing in Sweden (NAISS), funded by the Swedish Research Council through grant agreement no. 2022-06725.

\clearpage
\onecolumngrid
\section*{Supplementary Information}
\renewcommand{\thefigure}{SI:\arabic{figure}}
\renewcommand{\thetable}{SI:\arabic{table}}
\setcounter{figure}{0}
\setcounter{table}{0}

\section{Model description}
\label{SI:model-description}

In \moldstruct, the standard harmonic potential for 2-body bonded interactions is replaced by a Morse potential, which allows bonds to stretch and dissociate under Coulomb forces that develop as the molecule ionises.  Once the interatomic separation exceeds twice the equilibrium bond length, $r > 2b_0$, the Morse contribution is permanently removed for that bond, reflecting the physical irreversibility of bond rupture at such extensions.  A short-range repulsive $r^{-12}$ term is retained for broken bonds to prevent unphysical overlap of atoms.  Angle and dihedral interactions carry no such intrinsic dissociation mechanism and a harmonic angle potential would continue exerting a restoring force toward the equilibrium geometry even in a highly atomised molecule, which is unphysical.  All 3- and higher-body bonded potentials, such as angles, Urey-Bradley 1-3 springs, proper and improper dihedrals, and backbone CMAP corrections, are therefore scaled according to the mean ionisation state of the participating atoms, \begin{equation}     V' = \min\!\left(1,\max\!\left(1-\bar{z},0\right)\right)V, \end{equation} where $V$ is the unmodified force-field potential and $\bar{z}$ is the arithmetic mean charge of the atoms involved, computed at each time step.  Both forces and energies are scaled consistently.  This ensures that covalent angle and torsional constraints vanish continuously and smoothly as ionisation progresses, without introducing discontinuities in the force field.  This scaling is physically exact for hydrogen, which loses all bonding character upon single ionisation, but may be over-aggressive for heavy atoms, which retain multiple valence electrons and well-defined bond angles at $\bar{z} = 1$; a per-element correction to the scaling onset could therefore be warranted for heavy atoms.  No such correction was found to be necessary in the benchmarking of \moldstruct against experimental CEI data for 2-iodopyridine, where \moldstruct reproduced the measured momentum distributions without modification.  The Morse potential, by contrast, is not scaled by the ionisation state, because it already contains an intrinsic dissociation mechanism: as bonds stretch under Coulomb repulsion the restoring force naturally asymptotes to zero, and the explicit breaking threshold at $r > 2b_0$ captures the physical irreversibility of this process.  Scaling the Morse well depth would double-count the bond-breaking physics already embedded in the potential's shape.  For the Urey-Bradley 1-3 spring, an additional irreversible breaking threshold analogous to the Morse bond criterion is applied: once the 1-3 distance exceeds twice its equilibrium value, $r_{13} > 2r_{13}^0$, the spring is permanently zeroed, providing a two-level hierarchy of structural disintegration: bond rupture followed by loss of angular geometry.  Because GROMACS excludes 1-3 atom pairs (connected by two covalent bonds) from the non-bonded interaction list, those pairs carry no electrostatic repulsion in a standard simulation.  In the ionised regime this exclusion is unphysical; an explicit Coulomb repulsion between the terminal atoms of each angle is therefore added directly within the angle routines once both atoms carry positive charge, recovering the physically correct repulsion without altering the non-bonded exclusion list.  The conventional force-field scaling of 1-4 Coulomb interactions by a fixed factor $f_\text{QQ}$ (typically 0.5--0.83, depending on force field) is designed to avoid double-counting with bonded terms.  As ionisation proceeds those bonded contributions vanish, so the 1-4 electrostatic scaling is continuously interpolated toward its full unscaled value, \begin{equation}     f_\text{QQ}^\text{eff} = 1 - \left(1 - f_\text{QQ}\right)     \min\!\left(1,\max\!\left(1-\bar{z}_{14},0\right)\right), \end{equation} where $\bar{z}_{14}$ is the mean charge of the two 1-4 atoms.  At zero ionisation this recovers the original force-field value; at high ionisation it yields the unscaled Coulomb interaction.

\subsection*{Relation to standard GROMACS}  In standard GROMACS 4.5.4 all bonded interactions are evaluated with fixed parameters regardless of atomic charge.  Bonds never break; angle and torsional forces persist at full strength throughout the simulation; and the 1-3 and 1-4 pair-exclusion treatment is static.  The MolDStruct modifications replace this charge-independent framework with one that tracks the instantaneous ionisation state at every timestep, ensuring that the force field degrades in a physically motivated way as covalent structure is progressively destroyed during the Coulomb explosion.  No modification is made to non-bonded pair interactions beyond the 1-3 and 1-4 corrections noted above; long-range Coulomb and Lennard-Jones forces between non-bonded atoms are handled by the standard GROMACS engine.  

\subsection*{Force-field recommendation}  The modifications are most complete when used with the \textbf{CHARMM27} force field~\cite{mackerell_charmm_1998}.  CHARMM27 provides explicit Urey-Bradley 1-3 springs, which activate the second irreversible breaking threshold and the associated angle-energy scaling; and it provides the CMAP backbone correction, whose smooth ionisation-dependent removal eliminates a source of force discontinuity present in the earlier step-function implementation.  Together, these features give CHARMM27 a more complete physical treatment of the transition from intact peptide to a fully atomised fragment cloud than is available with AMBER-family force fields (AMBER99SB, AMBER03, etc.), which lack Urey-Bradley and CMAP terms and therefore leave those code paths inactive.  AMBER remains a valid choice when no CHARMM27 parameterisation exists for the molecule under study, as all other modifications described above apply equally to both.

\section{Model benchmarking}
\label{SI:benchmarkmethods}

\subsubsection{Suitability of Density Functional Theory (DFT) as a benchmark reference}
\label{SI:dft-suitability}

While DFT-based approaches present known limitations for XFEL ionisation conditions, including the use of pseudopotentials that exclude core electrons and approximate treatment of exchange and correlation, the validity of ground-state DFT for modelling fragmentation dynamics under such conditions has been assessed in Ref.~\citep{Aminosyror}. There, DFT (SIESTA) results were benchmarked against all-electron Hartree-Fock calculations and excited-state time-dependent DFT simulations for the same amino acid fragmentation problem. Since we used QM specifically to compare fragmentation patterns and the timescales on which bonds break, it is in this regime, rather than in general electronic structure properties, that the agreement between DFT and Hartree-Fock is relevant. The fragmentation patterns were found to be qualitatively similar across all methods, supporting the use of QM as a reliable reference for benchmarking \moldstruct in this context. QM has further been used as a reference across a range of X-ray fragmentation studies\cite{lassi2024}, including x-ray-induced fragmentation of iodine-doped DNA oligomers~\cite{ouassim_chain} and DNA backbone breakage by x-ray photoactivation~\cite{svensson2025radiotherapy}.

\subsubsection{Computational time}
\label{SI:computational-cost}

The wall-clock time per simulation was measured for all runs of both methods
(Table~\ref{tab:walltime}).
Summing the 20 stochastic charge configurations per charge state and initial
structure gives a total \moldstruct\ computational time of $48 \pm 6$\,s per
(structure, charge state) pair.
QM/SIESTA required $10.0 \pm 5.4$\,h per run (median 8.4\,h; range
3.1--31.7\,h, $N=230$), reflecting the varying number of SCF iterations
needed for convergence at different charge states.
The full QM/SIESTA campaign accumulated 2310\,h of wall time,
compared to 3.1\,h for the complete \moldstruct\ campaign (20 configurations
per pair).
If a single stochastic configuration per (structure, charge-state) pair
suffices, the equivalent \moldstruct\ campaign ($N=230$) completes in
${\sim}9$\,min, a speedup of ${\sim}15\,000\times$ over QM/SIESTA.

\begin{table}[h]
\centering
\begin{tabular}{lccc}
\hline
 & \moldstruct  (20 configs) & \moldstruct  (1 config) & QM/SIESTA \\
\hline
Mean wall time        & $48 \pm 6$\,s         & $2.4 \pm 0.5$\,s             & $10.0 \pm 5.4$\,h \\
Total (all runs)      & 3.1\,h \;($N{=}4600$)  & ${\sim}9$\,min \;($N{=}230$) & 2310\,h \;($N{=}230$) \\
Speedup vs QM/SIESTA  & ${\sim}750\times$     & ${\sim}15\,000\times$         & --- \\
\hline
\end{tabular}
\caption{Wall-clock time for simulations with \moldstruct and QM/SIESTA on the gradual ionisation of di-alanine. ; $N$ represents the number of runs.
\textit{20 configs}: all 20 stochastic configurations per (structure, charge-state)
pair summed.
\textit{1 config}: one configuration per pair (hypothetical minimum).
}
\label{tab:walltime}
\end{table}

\subsubsection{Simulation time and energy convergence}
\label{SI:energy-convergence}
\begin{figure}[htbp]
    \centering
    \includegraphics[width=0.46\textwidth]{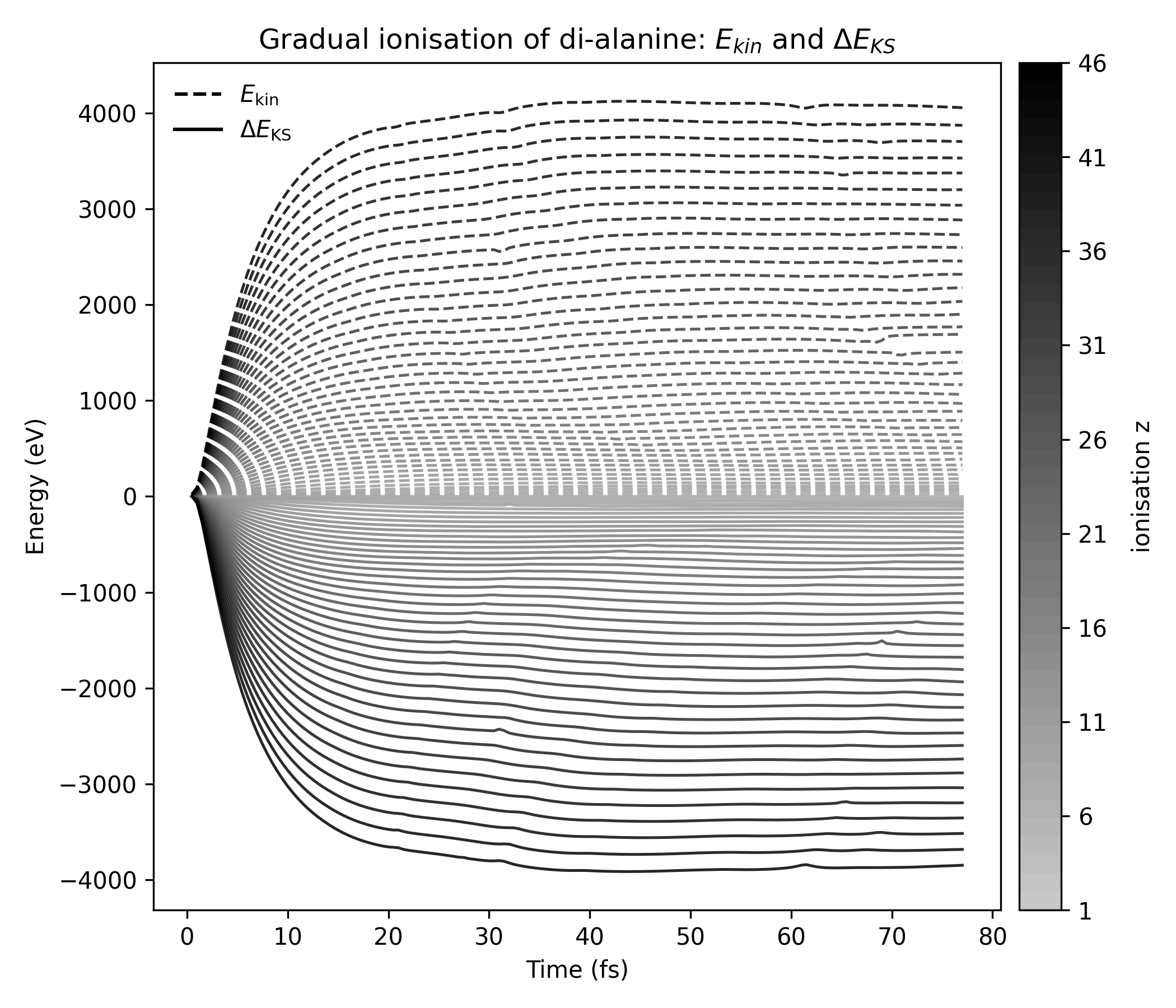}
    \hfill
    \includegraphics[width=0.51\textwidth]{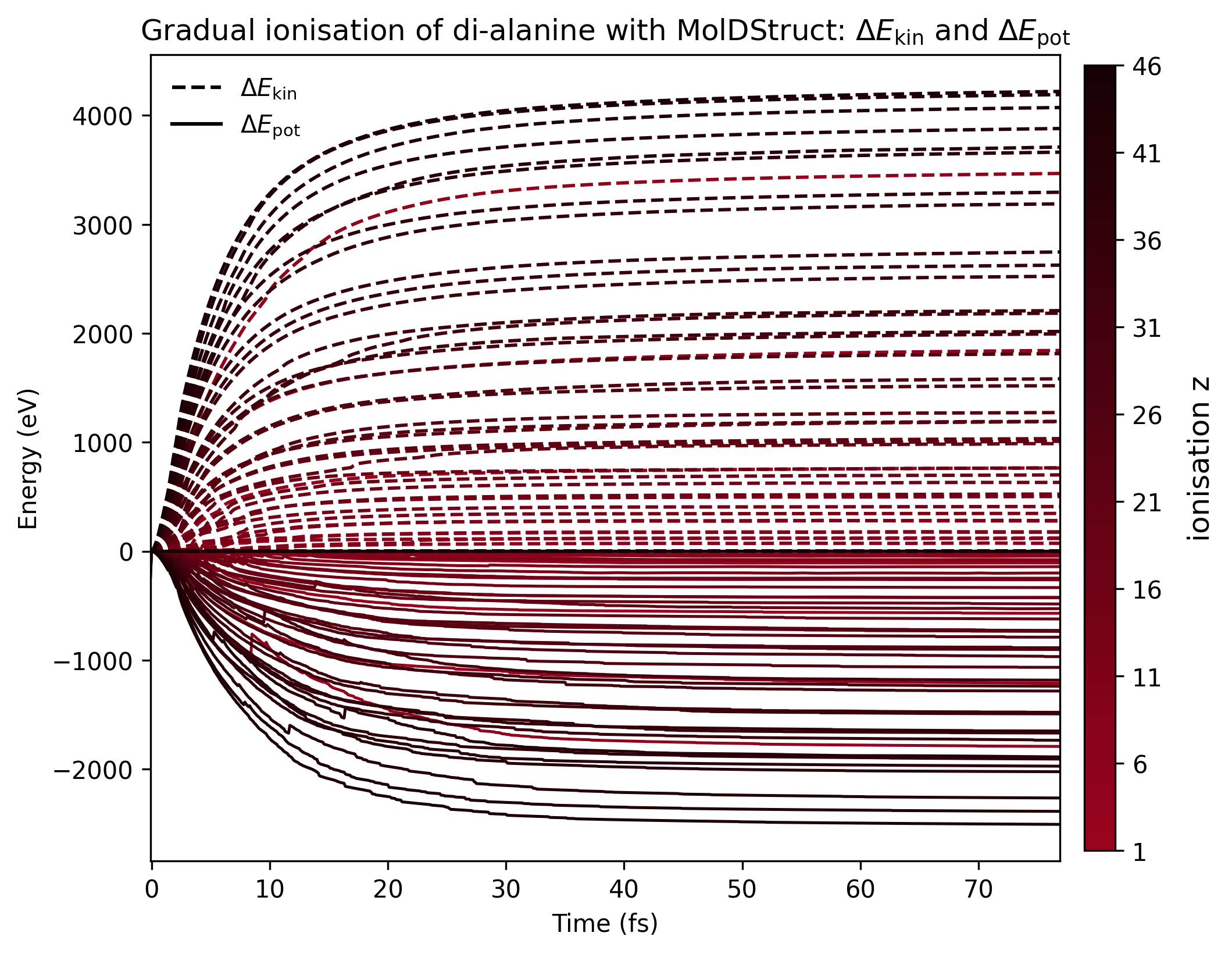}
    \caption{$\Delta E_\mathrm{kin}$ (dashed) and $\Delta E_\mathrm{pot}$ (solid) over 77~fs for one representative di-alanine structure ionised from $+1$ to $+46$, simulated with the QM/MD code (left, grey) and \moldstruct (right, red). Colour encodes ionisation state (bright: low, dark: high). Both panels plateau well before 77~fs, confirming the explosion is complete. In \moldstruct $|\Delta E_\mathrm{pot}|$ falls short of $|\Delta E_\mathrm{kin}|$ because each charge-update event injects Coulomb energy with no tracked source, which accumulates as excess kinetic energy.}
    \label{fig:2ala_siesta_energy} 
\end{figure}

~\autoref{fig:2ala_siesta_energy} shows the conversion of potential to kinetic energy during the simulations for QM/MD and \moldstruct and during the simulations. The kinetic energy plateaus before 77~fs, confirming the explosion is complete and the simulation time is sufficiently long.
MolDStruct computes $E_\mathrm{kin} = \tfrac{1}{2}\sum_i m_i v_i^2$ and reports
$E_\mathrm{pot}$ as the sum of Morse bond, short-range Lennard-Jones, short-range Coulomb
(from instantaneous integer charges), and bonded terms scaled by the ionisation factor
$f(\bar{z}) = \max(1-\bar{z},0)$; at the charge states considered, $E_\mathrm{Coul}$
dominates.  We plot $\Delta E = E(t)-E(0)$, so both curves start at zero.  Under energy
conservation $\Delta E_\mathrm{kin} = -\Delta E_\mathrm{pot}$, but in practice
$|\Delta E_\mathrm{kin}|$ exceeds $|\Delta E_\mathrm{pot}|$ because the charge-transfer
module (\texttt{userint2}) assigns additional charges during the explosion: each event
raises $E_\mathrm{Coul}$ discontinuously without a tracked energy source, injecting net
energy that accumulates in $\Delta E_\mathrm{kin}$.

\subsubsection{Effect of temperature on QM fragmentation patterns}
\label{SI:temperature}

\begin{figure}[!ht]
    \centering
    \includegraphics[width=0.7
    \linewidth]{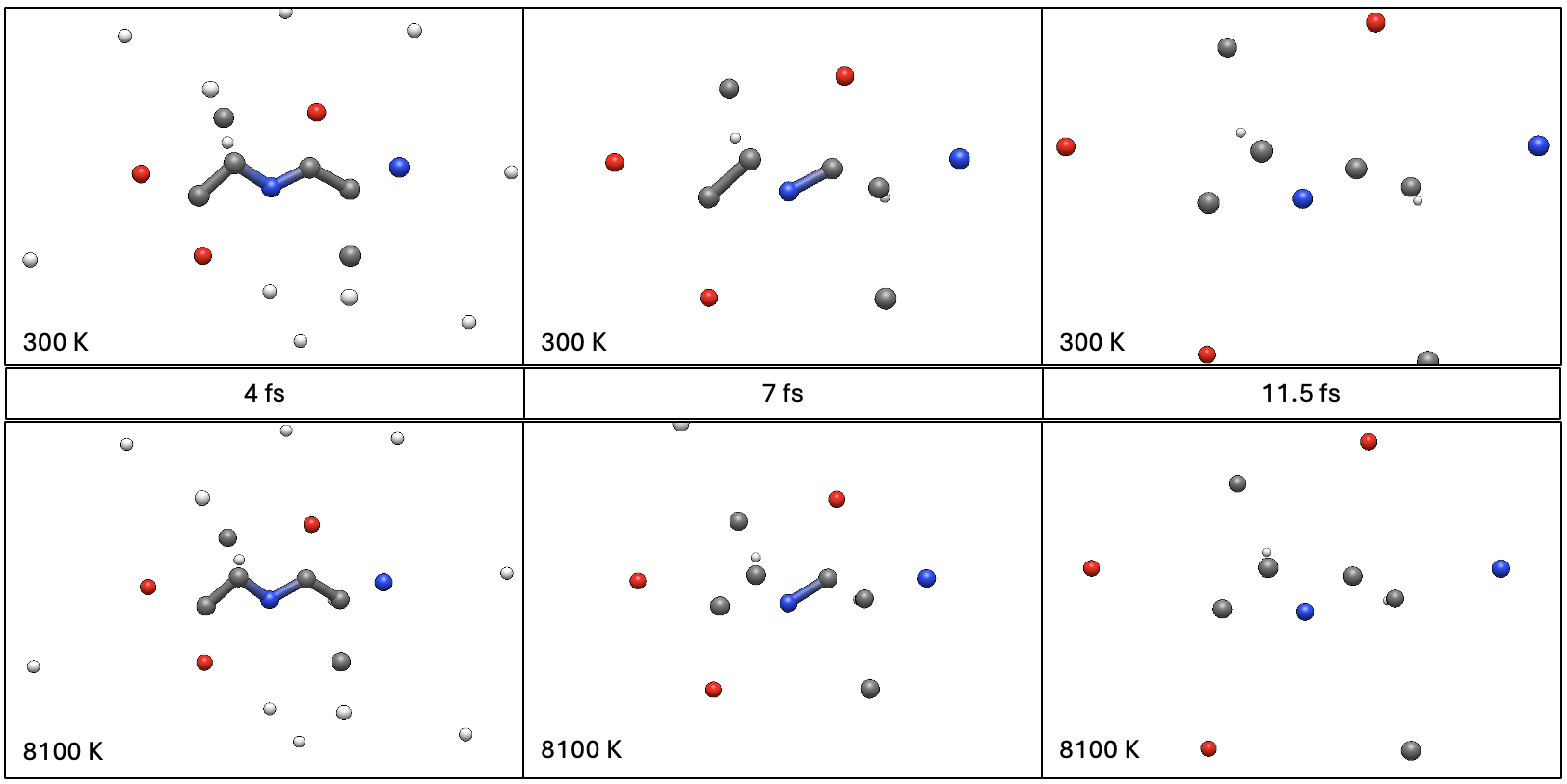}
    \caption{Comparison of Coulomb explosion of di-alanine ionised to $+46$ with the QM code. The upper row shows a simulation with electronic and initial temperature of 300~K, while the lower row shows a simulation at 8100~K. At 300~K the code converges only for the first 11.5--21~fs, whereas at 8100~K the full 77.5~fs can be simulated. Despite the temperature difference, the fragmentation patterns are similar. Atoms are colour-coded by element: C (grey), H (white), N (blue), O (red). Atom labels follow the convention of Fig.~\ref{fig:2Ala_structure}.}
    \label{fig:2Ala_temperature_siesta}
\end{figure}

~\autoref{fig:2Ala_temperature_siesta} illustrates the fragmentation of di-alanine ionised to a total charge of $+46$, as simulated using QM calculations at two different temperatures, 300~K and 8100~K, and at time steps of 4~fs, 7~fs, and 11.5~fs. For the simulation at 300~K, the QM code ceases to converge beyond these time steps, whereas at 8100~K the simulation remains stable and proceeds until the predefined end time of 77~fs. Despite the substantial difference in temperature, the resulting fragmentation patterns are remarkably similar. This observation indicates that fragmentation is primarily governed by the degree of ionisation rather than the thermal conditions, thereby justifying the use of elevated temperatures to improve numerical convergence of the QM simulations.

\subsubsection{Bond integrity and fragmentation patterns}
\label{SI:bond-integrity}

\begin{figure}[H]
    \centering
    \begin{minipage}[c]{0.40\textwidth}
        \centering
        \includegraphics[width=\textwidth]{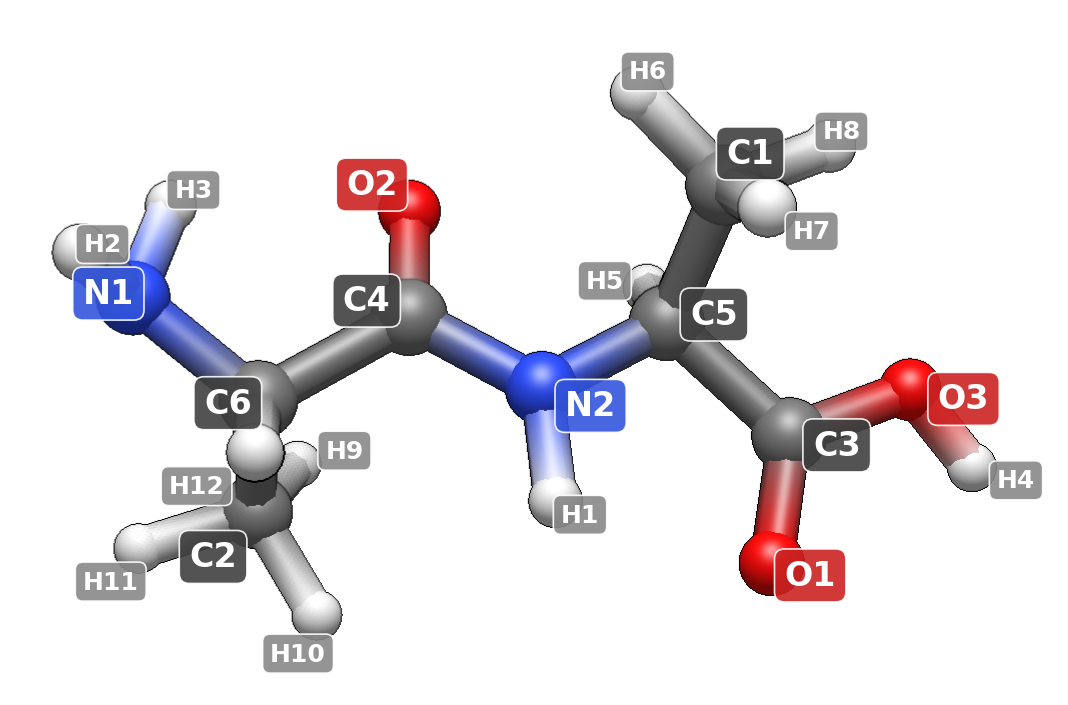}
    \end{minipage}%
    \hspace{1.5em}%
    \begin{minipage}[c]{0.38\textwidth}
        \caption{Di-alanine atom labelling used as reference for bond indices in Figs.~\ref{fig:2Ala_non_H_bond_integrity} and~\ref{fig:2Ala_H_bond_integrity}.}
        \label{fig:2Ala_structure}
    \end{minipage}
\end{figure}

~\autoref{fig:2Ala_structure} shows the chemical structure of di-alanine with labelled atoms.
The atom labels serve as reference for the bond indices in the bond integrity plots of Figs.~\ref{fig:2Ala_non_H_bond_integrity} and~\ref{fig:2Ala_H_bond_integrity}.

Data analysis and visualisation were performed in Python~\cite{Python} using the NumPy~\cite{Numpy}, pandas~\cite{pandas}, Matplotlib~\cite{Matplotlib}, and MDAnalysis~\cite{MDAnalysis} packages. The Souvatzis bond integrity parameter $\mathcal{B}$~\cite{bond_integrity} ranges from 0 to 1, with 1 indicating a fully intact bond and 0.5 or below indicating that the bond, on average, is broken. Bond-breaking events were recorded from the bond integrity trajectories to identify fragment species at the end of each simulation. Bond integrity values for QM were averaged across the five initial structures; for \moldstruct, results were first averaged over the 20 simulations per structure, then averaged across the 5 initial structures.

\begin{figure}[htbp]
    \centering
    \includegraphics[width=0.75\textwidth]{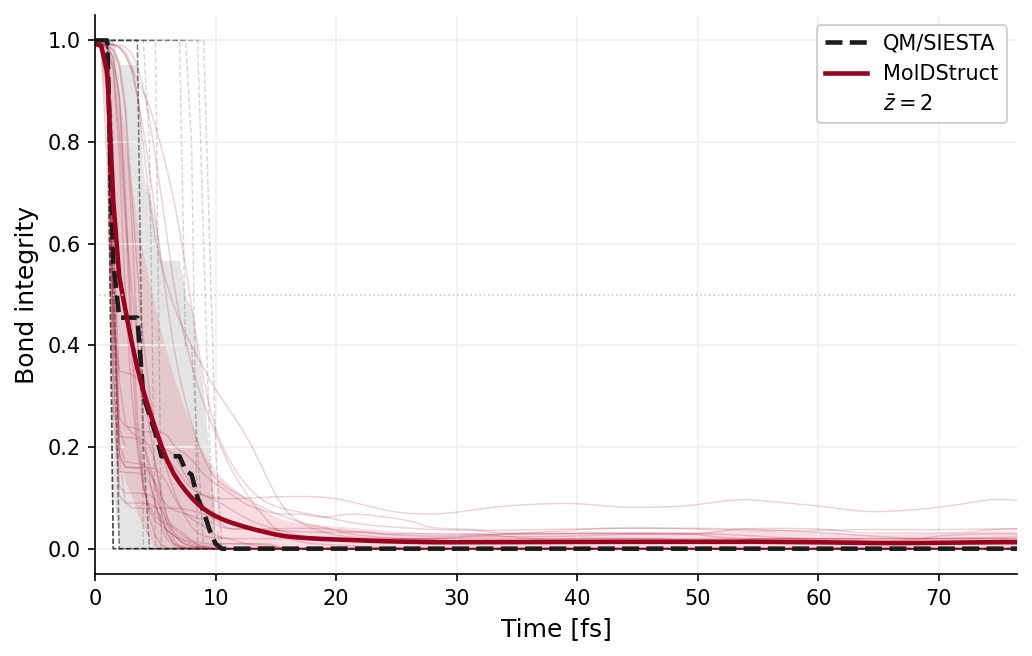}
    \caption{Mean bond integrity over time at $\bar{z}=2$ ($z=46$) for
    di-alanine, comparing QM/SIESTA (dashed, dark) and \moldstruct\
    (solid, red). Thick lines show the mean over all 22 bonds; shaded
    bands indicate the standard deviation across bonds; thin lines show
    individual bonds.}
    \label{fig:BI_zbar2}
\end{figure}

Figure~\ref{fig:BI_zbar2} shows the bond integrity as a function of time at $\bar{z}=2$ for di-alanine. Both methods predict near-complete bond dissociation within $\sim$20\,fs. The mean bond integrity curves are in close agreement: both show the same rapid decay and similar spread across bond types, confirming that classical Coulomb forces dominate fragmentation dynamics at full ionisation. The QM curve drops marginally faster in the first 5\,fs, reflecting the different initial conditions of the two methods. In QM/SIESTA, the simulation starts from a charge-relaxed geometry in which the electronic charge is already distributed over molecular orbitals, so Coulomb-driven bond stretching begins immediately. In \moldstruct, integer charges are assigned at $t=0$ and charge redistribution via the over-barrier model occurs at the start of the simulation, introducing a brief delay before the Coulomb explosion fully develops.

~\autoref{fig:2Ala_non_H_bond_integrity} shows the bond integrity of all non-hydrogen bonds in di-alanine as a function of ionisation degree (x-axis, mean charge per atom $\bar{z}$) and simulation time (y-axis, in fs), with QM/SIESTA results on the left (grey/black) and \moldstruct\ results on the right (red).
~\autoref{fig:2Ala_H_bond_integrity} shows the corresponding bond integrity plots for hydrogen bonds.
In both figures, intact bonds (bond integrity $> 0.5$) appear as coloured regions, while faded or white regions indicate broken bonds (bond integrity $< 0.5$).
Values are averaged over 5 initial structures; for \moldstruct, results are first averaged over 20 simulations per structure, then averaged across the 5 initial structures. At lower ionisation levels, where the two methods differed in their fragmentation patterns, differences in bond-breaking order and timing were also apparent, reflecting the fundamental discrepancy in charge treatment between the two methods.

The \moldstruct plots exhibit more intermediate bond integrity values, appearing as intermediate colours rather than the near-binary black-and-white transitions of the QM plots. Three factors contribute. First, \moldstruct results are averaged over 20 simulations per structure before averaging across the five initial structures, whereas QM results are averaged over five structures only; bonds that break in some runs but not others therefore produce intermediate mean values. Second, integer charge assignment distributes the total charge stochastically across atoms at $t=0$, generating a wider spread of initial conditions and bond-breaking outcomes across those 20 runs. Third, the over-barrier charge transfer module allows hydrogen atoms to redistribute charge to neighbouring acceptors at varying times across runs, further diversifying fragmentation pathways. In QM/SIESTA, by contrast, each structure carries a single continuously distributed charge state, so bond breaking proceeds from a well-defined initial condition, yielding near-binary transitions per structure.

\begin{figure}[H]
    \centering
    \includegraphics[width=0.48\textwidth]{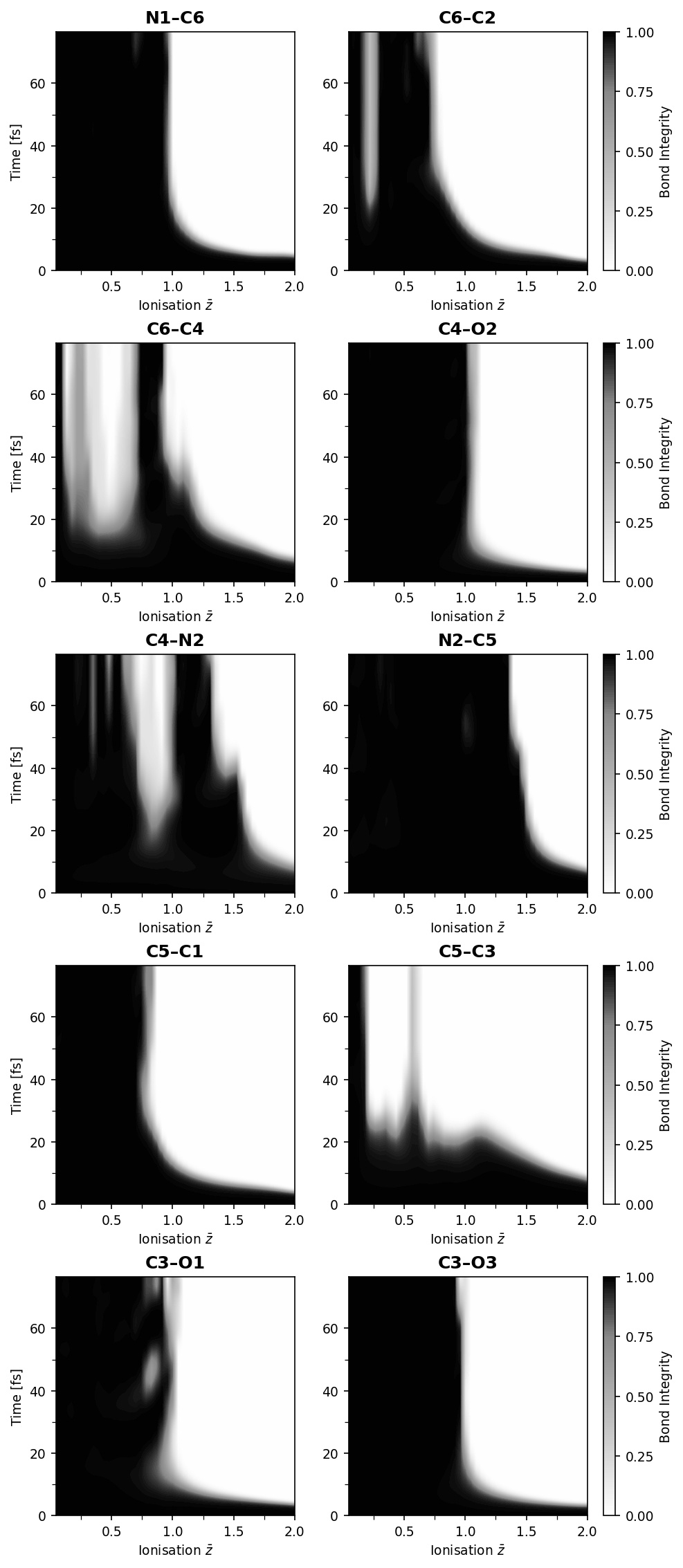}%
    \hfill%
    \includegraphics[width=0.48\textwidth]{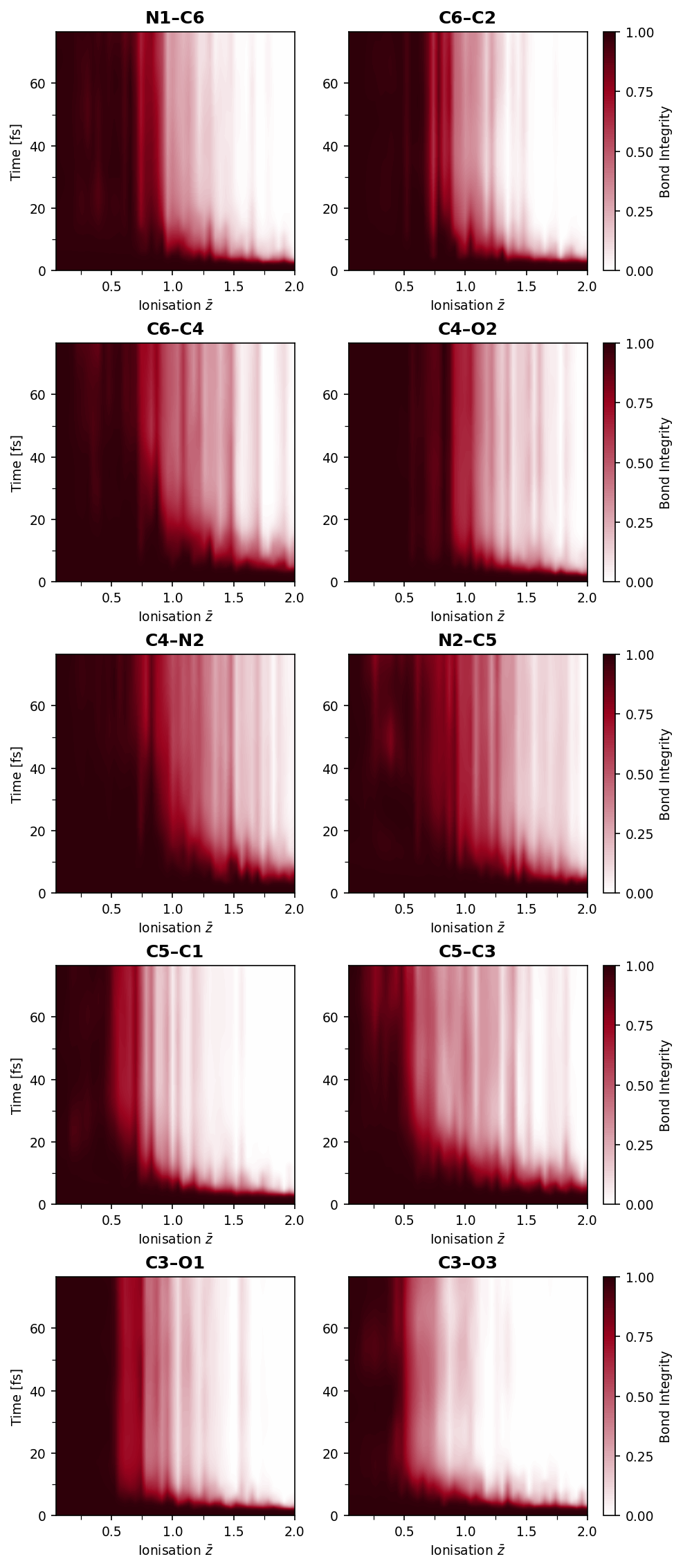}
    \caption{Bond integrity of non-hydrogen bonds in di-alanine as a function of ionisation degree (x-axis, mean charge per atom $\bar{z}$) and simulation time (y-axis, in fs). Left: QM/SIESTA simulations (grey/black). Right: \moldstruct simulations (red). Coloured regions indicate intact bonds (bond integrity $> 0.5$); faded/white regions indicate broken bonds ($< 0.5$). Values are averaged over 5 initial structures; \moldstruct results are additionally averaged over 20 simulations per structure.}
    \label{fig:2Ala_non_H_bond_integrity}
\end{figure}

\begin{figure}[H]
    \centering
    \includegraphics[width=0.48\textwidth]{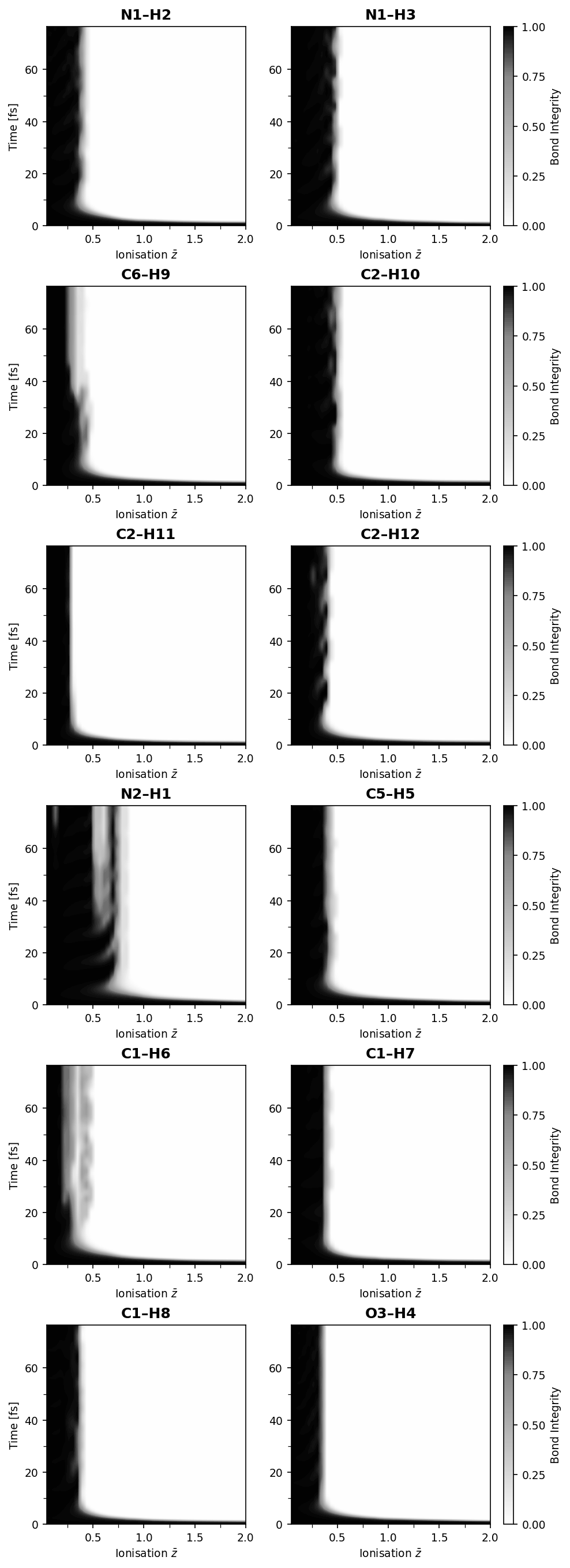}%
    \hfill%
    \includegraphics[width=0.48\textwidth]{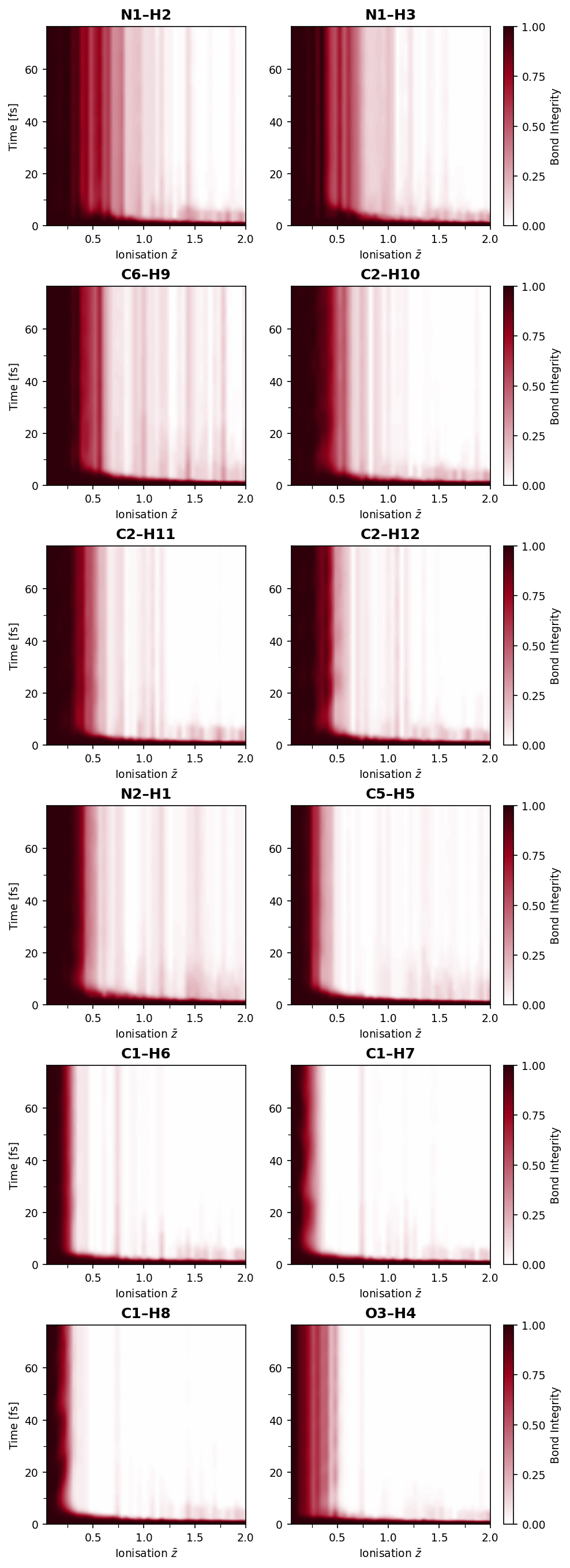}
    \caption{Bond integrity of hydrogen bonds in di-alanine as a function of ionisation degree (x-axis, mean charge per atom $\bar{z}$) and simulation time (y-axis, in fs). Left: QM/SIESTA simulations (grey/black). Right: \moldstruct simulations (red). Coloured regions indicate intact bonds (bond integrity $> 0.5$); faded/white regions indicate broken bonds ($< 0.5$). Values are averaged over 5 initial structures; \moldstruct results are additionally averaged over 20 simulations per structure.}
    \label{fig:2Ala_H_bond_integrity}
\end{figure}

\subsubsection{Fragment species}
\label{SI:fragments}

~\autoref{tab:diala-frags} lists the chemical formulas, masses, and occurrence counts of the fragment species obtained from QM simulations and MolDStruct calculations of gradually ionised di-alanine, corresponding to Fig.~3 of the main text.

% In the transition regime between $\bar{z} = 1.13$ and $\bar{z} = 1.35$, matching fragments were observed for small molecular species, including $\mathrm{CO}$, $\mathrm{C_2}$, $\mathrm{C_2N}$, and $\mathrm{C_2NO}$.

%\clearpage
%\newpage
\begin{center}
{\footnotesize
\begin{longtable}{c|c|p{6.5cm}|p{6.5cm}}
\caption{List of the chemical formulas, masses, and occurrence counts of the fragment species obtained from QM (left) and MolDStruct calculations (right) of gradually ionised di-alanine which are shown in Figure~2 of the main paper.} \label{tab:diala-frags} \\
\hline
\multicolumn{2}{c|}{\textbf{Ionisation}} & \textbf{QM Code} & \textbf{MolDStruct} \\
\cline{1-2}
$\bar{z}$ & $z$
& \textbf{(Fragment, Mass [u], Count)}
& \textbf{(Fragment, Mass [u], Count)} \\
\hline
\endfirsthead

\hline
\multicolumn{2}{c|}{\textbf{Ionisation}} & \textbf{QM Code} & \textbf{MolDStruct} \\
\cline{1-2}
$\bar{z}$ & $z$
& \textbf{(Fragment, Mass [u], Count)}
& \textbf{(Fragment, Mass [u], Count)} \\
\hline
\endhead

\hline
\endfoot
0.04 & 1 & ($\mathrm{C_{6}H_{12}N_{2}O_{3}}$, 160u, $1$) & ($\mathrm{C_{6}H_{12}N_{2}O_{3}}$, 160u, $1$) \\
0.09 & 2 & ($\mathrm{C_{4}H_{6}NO_{3}}$, 116u, $1$)$ | $($\mathrm{C_{2}H_{6}N}$, 44u, $1$) & ($\mathrm{C_{6}H_{12}N_{2}O_{3}}$, 160u, $1$) \\
0.13 & 3 & ($\mathrm{C_{4}H_{6}NO_{3}}$, 116u, $1$)$ | $($\mathrm{C_{2}H_{6}N}$, 44u, $1$) & ($\mathrm{C_{6}H_{12}N_{2}O_{3}}$, 160u, $1$) \\
0.17 & 4 & ($\mathrm{C_{3}H_{5}NO}$, 71u, $1$)$ | $($\mathrm{CH_{3}}$, 15u, $1$)$ | $($\mathrm{CH_{3}N}$, 29u, $1$)$ | $($\mathrm{CHO_{2}}$, 45u, $1$) & ($\mathrm{C_{6}H_{12}N_{2}O_{3}}$, 160u, $1$) \\
0.22 & 5 & ($\mathrm{C_{3}H_{4}NO}$, 70u, $1$)$ | $($\mathrm{CH_{3}}$, 15u, $1$)$ | $($\mathrm{CH_{3}N}$, 29u, $1$)$ | $($\mathrm{CHO_{2}}$, 45u, $1$)$ | $($\mathrm{H}$, 1u, $1$) & ($\mathrm{C_{6}H_{12}N_{2}O_{3}}$, 160u, $1$) \\
0.26 & 6 & ($\mathrm{C_{3}H_{4}NO}$, 70u, $1$)$ | $($\mathrm{C_{2}H_{5}N}$, 43u, $1$)$ | $($\mathrm{CHO_{2}}$, 45u, $1$)$ | $($\mathrm{H}$, 1u, $2$) & ($\mathrm{C_{6}H_{10}N_{2}O_{3}}$, 158u, $1$)$ | $($\mathrm{H}$, 1u, $2$) \\
0.30 & 7 & ($\mathrm{C_{3}H_{4}NO}$, 70u, $1$)$ | $($\mathrm{C_{2}H_{4}N}$, 42u, $1$)$ | $($\mathrm{CO_{2}}$, 44u, $1$)$ | $($\mathrm{H}$, 1u, $4$) & ($\mathrm{C_{6}H_{9}N_{2}O_{3}}$, 157u, $1$)$ | $($\mathrm{H}$, 1u, $3$) \\
0.35 & 8 & ($\mathrm{C_{3}H_{2}NO}$, 68u, $1$)$ | $($\mathrm{C_{2}H_{3}N}$, 41u, $1$)$ | $($\mathrm{CO_{2}}$, 44u, $1$)$ | $($\mathrm{H}$, 1u, $7$) & ($\mathrm{C_{6}H_{7}N_{2}O_{3}}$, 155u, $1$)$ | $($\mathrm{H}$, 1u, $5$) \\
0.39 & 9 & ($\mathrm{C_{3}HNO}$, 67u, $1$)$ | $($\mathrm{C_{2}H_{2}N}$, 40u, $1$)$ | $($\mathrm{CO_{2}}$, 44u, $1$)$ | $($\mathrm{H}$, 1u, $9$) & ($\mathrm{C_{6}H_{7}N_{2}O_{3}}$, 155u, $1$)$ | $($\mathrm{H}$, 1u, $5$) \\
0.43 & 10 & ($\mathrm{C_{3}HNO}$, 67u, $1$)$ | $($\mathrm{C_{2}H_{2}N}$, 40u, $1$)$ | $($\mathrm{CO_{2}}$, 44u, $1$)$ | $($\mathrm{H}$, 1u, $9$) & ($\mathrm{C_{6}H_{6}N_{2}O_{3}}$, 154u, $1$)$ | $($\mathrm{H}$, 1u, $6$) \\
0.48 & 11 & ($\mathrm{C_{3}NO}$, 66u, $1$)$ | $($\mathrm{C_{2}N}$, 38u, $1$)$ | $($\mathrm{CO_{2}}$, 44u, $1$)$ | $($\mathrm{H}$, 1u, $12$) & ($\mathrm{C_{6}H_{5}N_{2}O_{3}}$, 153u, $1$)$ | $($\mathrm{H}$, 1u, $7$) \\
0.52 & 12 & ($\mathrm{C_{3}NO}$, 66u, $1$)$ | $($\mathrm{C_{2}N}$, 38u, $1$)$ | $($\mathrm{CO_{2}}$, 44u, $1$)$ | $($\mathrm{H}$, 1u, $12$) & ($\mathrm{C_{6}H_{2}N_{2}O_{3}}$, 150u, $1$)$ | $($\mathrm{H}$, 1u, $10$) \\
0.57 & 13 & ($\mathrm{C_{3}NO}$, 66u, $1$)$ | $($\mathrm{C_{2}N}$, 38u, $1$)$ | $($\mathrm{CO_{2}}$, 44u, $1$)$ | $($\mathrm{H}$, 1u, $12$) & ($\mathrm{C_{6}H_{2}N_{2}O_{3}}$, 150u, $1$)$ | $($\mathrm{H}$, 1u, $10$) \\
0.61 & 14 & ($\mathrm{C_{2}N}$, 38u, $2$)$ | $($\mathrm{CO_{2}}$, 44u, $1$)$ | $($\mathrm{CO}$, 28u, $1$)$ | $($\mathrm{H}$, 1u, $12$) & ($\mathrm{C_{5}H_{2}N_{2}O}$, 106u, $1$)$ | $($\mathrm{H}$, 1u, $10$)$ | $($\mathrm{CO}$, 28u, $1$)$ | $($\mathrm{O}$, 16u, $1$) \\
0.65 & 15 & ($\mathrm{C_{2}HN}$, 39u, $1$)$ | $($\mathrm{C_{3}NO}$, 66u, $1$)$ | $($\mathrm{CO_{2}}$, 44u, $1$)$ | $($\mathrm{H}$, 1u, $11$) & ($\mathrm{C_{6}HN_{2}O_{2}}$, 133u, $1$)$ | $($\mathrm{H}$, 1u, $11$)$ | $($\mathrm{O}$, 16u, $1$) \\
0.70 & 16 & ($\mathrm{C_{2}N}$, 38u, $1$)$ | $($\mathrm{CO_{2}}$, 44u, $1$)$ | $($\mathrm{C_{2}NO}$, 54u, $1$)$ | $($\mathrm{H}$, 1u, $12$)$ | $($\mathrm{C}$, 12u, $1$) & ($\mathrm{C_{6}N_{2}O_{2}}$, 132u, $1$)$ | $($\mathrm{H}$, 1u, $12$)$ | $($\mathrm{O}$, 16u, $1$) \\
0.74 & 17 & ($\mathrm{C_{2}N}$, 38u, $1$)$ | $($\mathrm{CO_{2}}$, 44u, $1$)$ | $($\mathrm{C_{2}NO}$, 54u, $1$)$ | $($\mathrm{H}$, 1u, $12$)$ | $($\mathrm{C}$, 12u, $1$) & ($\mathrm{C_{5}N_{2}O}$, 104u, $1$)$ | $($\mathrm{H}$, 1u, $12$)$ | $($\mathrm{CO}$, 28u, $1$)$ | $($\mathrm{O}$, 16u, $1$) \\
0.78 & 18 & ($\mathrm{CO_{2}}$, 44u, $1$)$ | $($\mathrm{C_{2}NO}$, 54u, $1$)$ | $($\mathrm{CN}$, 26u, $1$)$ | $($\mathrm{C}$, 12u, $2$)$ | $($\mathrm{H}$, 1u, $12$) & ($\mathrm{C_{4}N_{2}O}$, 92u, $1$)$ | $($\mathrm{H}$, 1u, $12$)$ | $($\mathrm{C}$, 12u, $2$)$ | $($\mathrm{O}$, 16u, $2$) \\
0.83 & 19 & ($\mathrm{CO_{2}}$, 44u, $1$)$ | $($\mathrm{C_{2}NO}$, 54u, $1$)$ | $($\mathrm{CN}$, 26u, $1$)$ | $($\mathrm{C}$, 12u, $2$)$ | $($\mathrm{H}$, 1u, $12$) & ($\mathrm{C_{4}N_{2}O}$, 92u, $1$)$ | $($\mathrm{H}$, 1u, $12$)$ | $($\mathrm{C}$, 12u, $2$)$ | $($\mathrm{O}$, 16u, $2$) \\
0.87 & 20 & ($\mathrm{CO_{2}}$, 44u, $1$)$ | $($\mathrm{C_{2}NO}$, 54u, $1$)$ | $($\mathrm{CN}$, 26u, $1$)$ | $($\mathrm{C}$, 12u, $2$)$ | $($\mathrm{H}$, 1u, $12$) & ($\mathrm{C_{4}N_{2}O}$, 92u, $1$)$ | $($\mathrm{H}$, 1u, $12$)$ | $($\mathrm{C}$, 12u, $2$)$ | $($\mathrm{O}$, 16u, $2$) \\
0.91 & 21 & ($\mathrm{CO}$, 28u, $1$)$ | $($\mathrm{CN}$, 26u, $2$)$ | $($\mathrm{C}$, 12u, $3$)$ | $($\mathrm{H}$, 1u, $12$)$ | $($\mathrm{O}$, 16u, $2$) & ($\mathrm{C_{3}N_{2}O}$, 80u, $1$)$ | $($\mathrm{H}$, 1u, $12$)$ | $($\mathrm{C}$, 12u, $3$)$ | $($\mathrm{O}$, 16u, $2$) \\
0.96 & 22 & ($\mathrm{CO}$, 28u, $1$)$ | $($\mathrm{C_{2}NO}$, 54u, $1$)$ | $($\mathrm{C}$, 12u, $3$)$ | $($\mathrm{H}$, 1u, $12$)$ | $($\mathrm{N}$, 14u, $1$)$ | $($\mathrm{O}$, 16u, $1$) & ($\mathrm{N}$, 14u, $1$)$ | $($\mathrm{H}$, 1u, $12$)$ | $($\mathrm{C_{3}NO}$, 66u, $1$)$ | $($\mathrm{C}$, 12u, $3$)$ | $($\mathrm{O}$, 16u, $2$) \\
1.00 & 23 & ($\mathrm{C_{2}N}$, 38u, $1$)$ | $($\mathrm{C}$, 12u, $4$)$ | $($\mathrm{H}$, 1u, $12$)$ | $($\mathrm{N}$, 14u, $1$)$ | $($\mathrm{O}$, 16u, $3$) & ($\mathrm{N}$, 14u, $1$)$ | $($\mathrm{H}$, 1u, $12$)$ | $($\mathrm{C}$, 12u, $4$)$ | $($\mathrm{C_{2}NO}$, 54u, $1$)$ | $($\mathrm{O}$, 16u, $2$) \\
1.04 & 24 & ($\mathrm{C_{2}N}$, 38u, $1$)$ | $($\mathrm{C}$, 12u, $4$)$ | $($\mathrm{H}$, 1u, $12$)$ | $($\mathrm{N}$, 14u, $1$)$ | $($\mathrm{O}$, 16u, $3$) & ($\mathrm{N}$, 14u, $1$)$ | $($\mathrm{H}$, 1u, $12$)$ | $($\mathrm{C}$, 12u, $4$)$ | $($\mathrm{C_{2}NO}$, 54u, $1$)$ | $($\mathrm{O}$, 16u, $2$) \\
1.09 & 25 & ($\mathrm{C_{2}N}$, 38u, $1$)$ | $($\mathrm{C}$, 12u, $4$)$ | $($\mathrm{H}$, 1u, $12$)$ | $($\mathrm{N}$, 14u, $1$)$ | $($\mathrm{O}$, 16u, $3$) & ($\mathrm{N}$, 14u, $1$)$ | $($\mathrm{H}$, 1u, $12$)$ | $($\mathrm{C_{2}NO}$, 54u, $1$)$ | $($\mathrm{C}$, 12u, $4$)$ | $($\mathrm{O}$, 16u, $2$) \\
1.13 & 26 & ($\mathrm{C_{2}N}$, 38u, $1$)$ | $($\mathrm{C}$, 12u, $4$)$ | $($\mathrm{H}$, 1u, $12$)$ | $($\mathrm{N}$, 14u, $1$)$ | $($\mathrm{O}$, 16u, $3$) & ($\mathrm{N}$, 14u, $1$)$ | $($\mathrm{H}$, 1u, $12$)$ | $($\mathrm{C}$, 12u, $4$)$ | $($\mathrm{C_{2}N}$, 38u, $1$)$ | $($\mathrm{O}$, 16u, $3$) \\
1.17 & 27 & ($\mathrm{C_{2}N}$, 38u, $1$)$ | $($\mathrm{C}$, 12u, $4$)$ | $($\mathrm{H}$, 1u, $12$)$ | $($\mathrm{N}$, 14u, $1$)$ | $($\mathrm{O}$, 16u, $3$) & ($\mathrm{N}$, 14u, $1$)$ | $($\mathrm{H}$, 1u, $12$)$ | $($\mathrm{C}$, 12u, $5$)$ | $($\mathrm{O}$, 16u, $3$)$ | $($\mathrm{CN}$, 26u, $1$) \\
1.22 & 28 & ($\mathrm{C_{2}N}$, 38u, $1$)$ | $($\mathrm{C}$, 12u, $4$)$ | $($\mathrm{H}$, 1u, $12$)$ | $($\mathrm{N}$, 14u, $1$)$ | $($\mathrm{O}$, 16u, $3$) & ($\mathrm{N}$, 14u, $1$)$ | $($\mathrm{H}$, 1u, $12$)$ | $($\mathrm{C}$, 12u, $5$)$ | $($\mathrm{CN}$, 26u, $1$)$ | $($\mathrm{O}$, 16u, $3$) \\
1.26 & 29 & ($\mathrm{C_{2}N}$, 38u, $1$)$ | $($\mathrm{C}$, 12u, $4$)$ | $($\mathrm{H}$, 1u, $12$)$ | $($\mathrm{N}$, 14u, $1$)$ | $($\mathrm{O}$, 16u, $3$) & ($\mathrm{N}$, 14u, $1$)$ | $($\mathrm{H}$, 1u, $12$)$ | $($\mathrm{C}$, 12u, $5$)$ | $($\mathrm{O}$, 16u, $3$)$ | $($\mathrm{CN}$, 26u, $1$) \\
1.30 & 30 & ($\mathrm{CN}$, 26u, $1$)$ | $($\mathrm{C}$, 12u, $5$)$ | $($\mathrm{H}$, 1u, $12$)$ | $($\mathrm{N}$, 14u, $1$)$ | $($\mathrm{O}$, 16u, $3$) & ($\mathrm{N}$, 14u, $2$)$ | $($\mathrm{H}$, 1u, $12$)$ | $($\mathrm{C}$, 12u, $6$)$ | $($\mathrm{O}$, 16u, $3$) \\
1.35 & 31 & ($\mathrm{C}$, 12u, $6$)$ | $($\mathrm{H}$, 1u, $12$)$ | $($\mathrm{N}$, 14u, $2$)$ | $($\mathrm{O}$, 16u, $3$) & ($\mathrm{N}$, 14u, $2$)$ | $($\mathrm{H}$, 1u, $12$)$ | $($\mathrm{C}$, 12u, $6$)$ | $($\mathrm{O}$, 16u, $3$) \\
1.39 & 32 & ($\mathrm{C}$, 12u, $6$)$ | $($\mathrm{H}$, 1u, $12$)$ | $($\mathrm{N}$, 14u, $2$)$ | $($\mathrm{O}$, 16u, $3$) & ($\mathrm{N}$, 14u, $2$)$ | $($\mathrm{H}$, 1u, $12$)$ | $($\mathrm{C}$, 12u, $6$)$ | $($\mathrm{O}$, 16u, $3$) \\
1.43 & 33 & ($\mathrm{C}$, 12u, $6$)$ | $($\mathrm{H}$, 1u, $12$)$ | $($\mathrm{N}$, 14u, $2$)$ | $($\mathrm{O}$, 16u, $3$) & ($\mathrm{N}$, 14u, $2$)$ | $($\mathrm{H}$, 1u, $12$)$ | $($\mathrm{C}$, 12u, $6$)$ | $($\mathrm{O}$, 16u, $3$) \\
1.48 & 34 & ($\mathrm{C}$, 12u, $6$)$ | $($\mathrm{H}$, 1u, $12$)$ | $($\mathrm{N}$, 14u, $2$)$ | $($\mathrm{O}$, 16u, $3$) & ($\mathrm{N}$, 14u, $2$)$ | $($\mathrm{H}$, 1u, $12$)$ | $($\mathrm{C}$, 12u, $6$)$ | $($\mathrm{O}$, 16u, $3$) \\
1.52 & 35 & ($\mathrm{C}$, 12u, $6$)$ | $($\mathrm{H}$, 1u, $12$)$ | $($\mathrm{N}$, 14u, $2$)$ | $($\mathrm{O}$, 16u, $3$) & ($\mathrm{N}$, 14u, $2$)$ | $($\mathrm{H}$, 1u, $12$)$ | $($\mathrm{C}$, 12u, $6$)$ | $($\mathrm{O}$, 16u, $3$) \\
1.57 & 36 & ($\mathrm{C}$, 12u, $6$)$ | $($\mathrm{H}$, 1u, $12$)$ | $($\mathrm{N}$, 14u, $2$)$ | $($\mathrm{O}$, 16u, $3$) & ($\mathrm{N}$, 14u, $2$)$ | $($\mathrm{H}$, 1u, $12$)$ | $($\mathrm{C}$, 12u, $6$)$ | $($\mathrm{O}$, 16u, $3$) \\
1.61 & 37 & ($\mathrm{C}$, 12u, $6$)$ | $($\mathrm{H}$, 1u, $12$)$ | $($\mathrm{N}$, 14u, $2$)$ | $($\mathrm{O}$, 16u, $3$) & ($\mathrm{N}$, 14u, $2$)$ | $($\mathrm{H}$, 1u, $12$)$ | $($\mathrm{C}$, 12u, $6$)$ | $($\mathrm{O}$, 16u, $3$) \\
1.65 & 38 & ($\mathrm{C}$, 12u, $6$)$ | $($\mathrm{H}$, 1u, $12$)$ | $($\mathrm{N}$, 14u, $2$)$ | $($\mathrm{O}$, 16u, $3$) & ($\mathrm{N}$, 14u, $2$)$ | $($\mathrm{H}$, 1u, $12$)$ | $($\mathrm{C}$, 12u, $6$)$ | $($\mathrm{O}$, 16u, $3$) \\
1.70 & 39 & ($\mathrm{C}$, 12u, $6$)$ | $($\mathrm{H}$, 1u, $12$)$ | $($\mathrm{N}$, 14u, $2$)$ | $($\mathrm{O}$, 16u, $3$) & ($\mathrm{N}$, 14u, $2$)$ | $($\mathrm{H}$, 1u, $12$)$ | $($\mathrm{C}$, 12u, $6$)$ | $($\mathrm{O}$, 16u, $3$) \\
1.74 & 40 & ($\mathrm{C}$, 12u, $6$)$ | $($\mathrm{H}$, 1u, $12$)$ | $($\mathrm{N}$, 14u, $2$)$ | $($\mathrm{O}$, 16u, $3$) & ($\mathrm{N}$, 14u, $2$)$ | $($\mathrm{H}$, 1u, $12$)$ | $($\mathrm{C}$, 12u, $6$)$ | $($\mathrm{O}$, 16u, $3$) \\
1.78 & 41 & ($\mathrm{C}$, 12u, $6$)$ | $($\mathrm{H}$, 1u, $12$)$ | $($\mathrm{N}$, 14u, $2$)$ | $($\mathrm{O}$, 16u, $3$) & ($\mathrm{N}$, 14u, $2$)$ | $($\mathrm{H}$, 1u, $12$)$ | $($\mathrm{C}$, 12u, $6$)$ | $($\mathrm{O}$, 16u, $3$) \\
1.83 & 42 & ($\mathrm{C}$, 12u, $6$)$ | $($\mathrm{H}$, 1u, $12$)$ | $($\mathrm{N}$, 14u, $2$)$ | $($\mathrm{O}$, 16u, $3$) & ($\mathrm{N}$, 14u, $2$)$ | $($\mathrm{H}$, 1u, $12$)$ | $($\mathrm{C}$, 12u, $6$)$ | $($\mathrm{O}$, 16u, $3$) \\
1.87 & 43 & ($\mathrm{C}$, 12u, $6$)$ | $($\mathrm{H}$, 1u, $12$)$ | $($\mathrm{N}$, 14u, $2$)$ | $($\mathrm{O}$, 16u, $3$) & ($\mathrm{N}$, 14u, $2$)$ | $($\mathrm{H}$, 1u, $12$)$ | $($\mathrm{C}$, 12u, $6$)$ | $($\mathrm{O}$, 16u, $3$) \\
1.91 & 44 & ($\mathrm{C}$, 12u, $6$)$ | $($\mathrm{H}$, 1u, $12$)$ | $($\mathrm{N}$, 14u, $2$)$ | $($\mathrm{O}$, 16u, $3$) & ($\mathrm{N}$, 14u, $2$)$ | $($\mathrm{H}$, 1u, $12$)$ | $($\mathrm{C}$, 12u, $6$)$ | $($\mathrm{O}$, 16u, $3$) \\
1.96 & 45 & ($\mathrm{C}$, 12u, $6$)$ | $($\mathrm{H}$, 1u, $12$)$ | $($\mathrm{N}$, 14u, $2$)$ | $($\mathrm{O}$, 16u, $3$) & ($\mathrm{N}$, 14u, $2$)$ | $($\mathrm{H}$, 1u, $12$)$ | $($\mathrm{C}$, 12u, $6$)$ | $($\mathrm{O}$, 16u, $3$) \\
2.00 & 46 & ($\mathrm{C}$, 12u, $6$)$ | $($\mathrm{H}$, 1u, $12$)$ | $($\mathrm{N}$, 14u, $2$)$ | $($\mathrm{O}$, 16u, $3$) & ($\mathrm{N}$, 14u, $2$)$ | $($\mathrm{H}$, 1u, $12$)$ | $($\mathrm{C}$, 12u, $6$)$ | $($\mathrm{O}$, 16u, $3$) \\
\end{longtable}
}
\end{center}

\subsubsection{Ion velocity and angular distributions: \moldstruct vs QM}
\label{SI:velocity-angular}

For \moldstruct, atomic velocities at the final simulation frame were read directly from the GROMACS output in nm\,ps$^{-1}$. For QM/SIESTA, velocities were not written directly; they were derived from atomic positions using a second-order central finite difference over the last three trajectory frames, $\mathbf{v}_i = (\mathbf{r}_i^{(N)} - \mathbf{r}_i^{(N-2)})/(2\Delta t)$, with $\Delta t = 0.5$\,fs, and converted from \AA\,fs$^{-1}$ to nm\,ps$^{-1}$. In both cases, the centre-of-mass velocity was subtracted so that reported velocities $\mathbf{v}_i^*$ reflect only Coulomb-driven fragment motion. The scalar speed is the magnitude of the corrected velocity vector; the polar angle is defined as $\cos\theta = v_z^*/|\mathbf{v}^*|$ and the azimuthal angle as $\varphi = \arctan2(v_y^*, v_x^*)$. Distributions were estimated using a kernel density estimator (KDE) with Silverman's bandwidth rule~\cite{silverman1986}. For \moldstruct, the 20 independent simulations per initial structure were averaged per structure before computing the KDE, yielding the same effective sample size as QM (5 structures, 23 atoms each).

\begin{figure}[htbp]
    \centering
    \includegraphics[width=\textwidth]{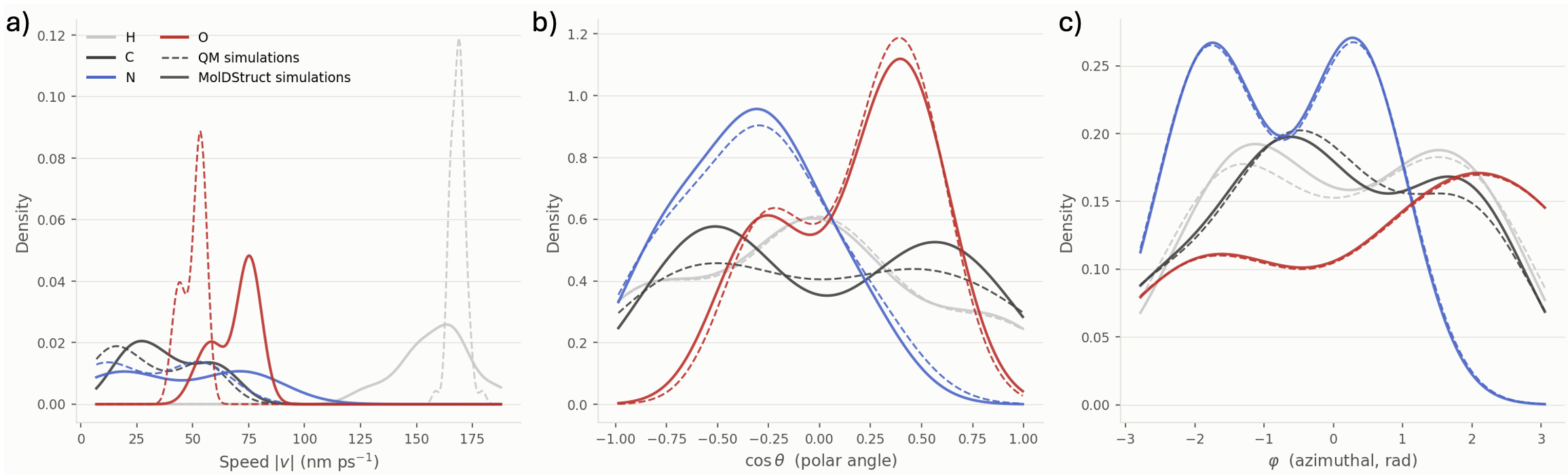}
\caption{Ion velocity and angular distributions for di-alanine at charge
state $\bar{z} \approx 2$ ($z = 46$), comparing QM/SIESTA (dashed) and
\moldstruct (solid) simulations. Colours indicate element: H (grey), C
(black), N (blue), O (red).
\textbf{(a)} Speed distributions: \moldstruct and QM agree for H,
C, and N; Oxygen shows an overestimate by \moldstruct relative to QM.
\textbf{(b)} Polar angle ($\cos\theta$) distributions: both methods are in
close agreement across all elements.
\textbf{(c)} Azimuthal angle ($\varphi$) distributions: similarly close
agreement between methods for all elements. The differences observed are
consistent with the use of static DFT rather than time-dependent DFT as
the QM reference~\cite{taylor2025}.}
\label{fig:2Ala_speed_angular}
\end{figure}

Figure~\ref{fig:2Ala_speed_angular} compares the speed and angular distributions of all four fragment species at $\bar{z} \approx 2$ which are H C N and O. The angular distributions in (b) and (c) are close between the two methods for all elements: both the polar ($\cos\theta$) and azimuthal ($\varphi$) projections show the same peak positions and widths for H, C, N, and Oxygen, indicating that \moldstruct captures the directional structure of the Coulomb explosion.  The speed distributions in (a) broadly follow the same pattern for H, C, and N, where the two methods produce overlapping distributions with no systematic offset; the hydrogen speed distribution is nevertheless broader than the DFT reference. Oxygen is the exception: \moldstruct predicts a median speed of $\approx 73$\,nm\,ps$^{-1}$ against $\approx 53$\,nm\,ps$^{-1}$ for QM, an overestimate of $\approx 38\%$.  Taylor et al.\ \cite{taylor2025} report that classical models overestimate the final kinetic energies of all ion species relative to TDDFT simulations. Our comparison shows this effect only for Oxygen, and not for H, C, or N. The key distinction is that Taylor's QM reference is time-dependent DFT (TDDFT), in which the laser field drives ionisation in real time and atomic charge states evolve gradually throughout the pulse. Our QM reference is SIESTA, a static DFT code where charges are fixed at their post-ionisation values from $t = 0$, the same assumption made in the classical model. Gradual charge build-up during the laser pulse is therefore absent from both the SIESTA and \moldstruct simulations.

The origins of these deviations are not directly tested in the present work; the following paragraphs propose possible explanations based on known model assumptions.

A possible explanation for the oxygen overestimate is that charge transfer in \moldstruct is implemented only for hydrogen as the electron donor, so oxygen atoms not bonded to hydrogen have no mechanism to redistribute positive charge to neighbouring atoms. In DFT, molecular orbital delocalisation distributes electron density continuously across the molecule, partially reducing the charge localised on individual oxygen atoms. In \moldstruct, oxygen may therefore accumulate more ionisation charge from photoionisation and Auger-Meitner decay than DFT predicts, driving stronger Coulomb repulsion and higher kinetic energy. A smaller systematic offset is also visible for carbon and nitrogen, though the speed distributions largely overlap.

A possible explanation for the broader hydrogen speed distribution lies in the stochastic charge assignment in \texttt{MolDStruct}. At $t = 0$, the total charge $z$ is distributed as integer charges across the 23 atoms.  The integer charge of hydrogen varies between runs, as does the charge on its neighbours. These differences in the initial charge configuration produce different Coulomb forces on hydrogen from the very start of the explosion. During the dynamics, the over-barrier model additionally allows hydrogen atoms carrying charge +1 to transfer that charge to a neighbouring acceptor when the local electrostatic potential suppresses the binding barrier; the timing of this event depends on the evolving local field, which itself reflects the varying initial charge distribution. In DFT/SIESTA, by contrast, charge is not integer but continuously distributed through the molecular orbital electron density, so hydrogen carries a well-defined fractional charge in every simulation. The discrete, stochastic character of the initial charge assignment in \texttt{MolDStruct}, and the subsequent over-barrier transfer events it conditions, may therefore account for the broader hydrogen speed distribution observed relative to the DFT reference.

\subsubsection{Gradual Ionisation of Alanine with MolDStruct and QM}
\label{supp:gradualIon}

\begin{figure}[htbp]
 \includegraphics[width=1\textwidth]{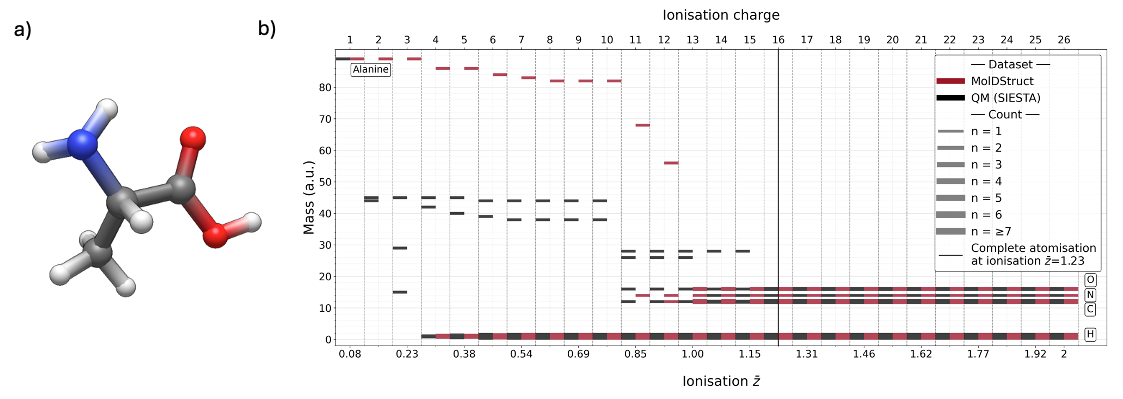}
    \caption{\textbf{a)} Atomic structure of alanine consisting of 13 atoms and without protonations. \textbf{b)} Fragmentation of gradually ionised alanine, simulated with the QM code (black) and MolDStruct (red). Colour darkness encodes fragment count (darker = more, up to 7). The y-axis shows fragment mass, from intact alanine (89\,u) down to single atoms H (1\,u), C (12\,u), N (14\,u), O (16\,u).}
    \label{fig:1Ala_combined}
\end{figure}

Additionally to di-alanine, we recorded the gradual ionisation of alanine from charge +1 to charge +26 to reach a maximum mean charge of 2 per atom. The structure of alanine has been modelled with the software avogadro and a model is shown in ~\autoref{fig:1Ala_combined} a).  Each simulation has been evolved for 77 fs with initial and electronic temperature set to 300K. For higher ionisation levels ($\bar{z}>0.5$), the temperature is increased with increasing ionisation, ranging from 600 K up to 6000 K for alanine, $\bar{z}=1.69$. From the atomic distances and the resulting bond integrities the fragments of the molecules have been recorded at the end of the simulations. The data obtained from MolDStruct and QM data is plotted in ~\autoref{fig:1Ala_combined} b). The y-axis shows the mass in atomic units. The x-axis records the degree of ionisation in full charges (top) and average charge per atom (bottom). At ionisation state 17, $\bar{z}=1.31$ the molecule explodes in single atoms for both codes, MolDStruct and QM. From this charge state on, both computation methods lead to the same fragments, confirming the $\bar{z}=1.35$ threshold observed with di-alanine for which \moldstruct and QM simulations agree.

\subsection{Comparison to experimental Data}
\label{2iod}

\subsubsection*{Newton plot construction}
Here, we describe the calculation of the momenta plotted in Fig.~4 of the main text from the \moldstruct simulations. The Newton plots display three-ion coincidences from sets of four measured ions; recording up to eightfold ion coincidences at the high repetition rate of the European XFEL yields identical-looking images~\cite{boll_x-ray_2022}, confirming that the three-ion coincidences capture the essential features of the asymptotic momentum distribution. Atomic velocities $\mathbf{v}_i(t)$ are stored in the GROMACS TRR format in units of nm\,ps$^{-1}$ and are read via the Python library MDAnalysis \cite{MDAnalysis}, which converts them internally to \AA\,ps$^{-1}$. The linear momentum of atom $i$ at time $t$ is
\begin{equation}
    \mathbf{p}_i(t) = m_i\,\mathbf{v}_i(t),
    \qquad [m_i]=\text{Da},\quad [\mathbf{v}_i]=\text{\AA\,ps}^{-1}.
\end{equation}

In accordance with the plots 1 b) and 1 d) in \citep{boll_x-ray_2022}, the momenta are converted into atomic units. One Da\,\AA\,ps$^{-1}$ corresponds to $1.66054\times10^{-25}$\,kg\,m\,s$^{-1}$. The atomic unit of momentum is $\hbar/a_0 = 1.99285\times10^{-24}$\,kg\,m\,s$^{-1}$, where $a_0$ is the Bohr radius. The conversion factor is therefore
\begin{equation}
    1\;\text{Da\,\AA\,ps}^{-1}
    = \frac{1.66054\times10^{-25}}{1.99285\times10^{-24}}\;\text{a.u.}
    \approx 0.08334\;\text{a.u.},
\end{equation}

so all momenta are obtained as $\mathbf{p}_i^{\,\mathrm{au}} = 0.08334\; m_i\,\mathbf{v}_i$. \\

The molecular Newton-plot frame is defined event-by-event from the asymptotic momenta at $t = 300$\,fs. Like in the experiment, the $z$-axis is chosen along the iodine recoil direction,

\begin{equation}
    \hat{z} = \frac{\mathbf{p}_{\mathrm{I}}}{\lvert\mathbf{p}_{\mathrm{I}}\rvert},
\end{equation}

and the $y$-axis is set to the component of the nitrogen momentum perpendicular
to $\hat{z}$, normalised to unit length,
\begin{equation}
    \hat{y} = \frac{\mathbf{p}_{\mathrm{N}}
                    - (\mathbf{p}_{\mathrm{N}}\cdot\hat{z})\,\hat{z}}
                   {\lvert\mathbf{p}_{\mathrm{N}}
                    - (\mathbf{p}_{\mathrm{N}}\cdot\hat{z})\,\hat{z}\rvert},
\end{equation}

ensuring $p_{\mathrm{N},y} > 0$ by construction.
The $x$-axis completes a right-handed system, $\hat{x} = \hat{y}\times\hat{z}$.
Each atomic momentum is then projected as
$p_{i,z} = \mathbf{p}_i\cdot\hat{z}$ and $p_{i,y} = \mathbf{p}_i\cdot\hat{y}$.

At $t = 300$\,fs, iodine, the heaviest fragment, carries the largest momentum ($\lvert\mathbf{p}_{\mathrm{I}}\rvert \approx 287$\,a.u.) directed along $+z$ by construction. Nitrogen acquires $(p_z,\,p_y)\approx(93,\,196)$\,a.u., placing it in the upper-right quadrant. The ring carbons are distributed around the full azimuth with magnitudes of $50$--$155$\,a.u.: C5 (diametrically opposite to iodine) reaches $p_z \approx -153$\,a.u., and C3 acquires $(p_z,\,p_y)\approx(74,-148)$\,a.u. Hydrogen atoms, the lightest fragments, attain $\lvert\mathbf{p}_{\mathrm{H}}\rvert \approx 71$\,a.u. The plot axes span $\pm 310$\,a.u.\ to accommodate the iodine star and the nitrogen cluster; tick marks are placed at $\pm 200$ and $\pm 100$\,a.u.

\subsubsection*{Explosion time evolution}
\begin{figure}
\centering
\includegraphics[width=0.7\textwidth]{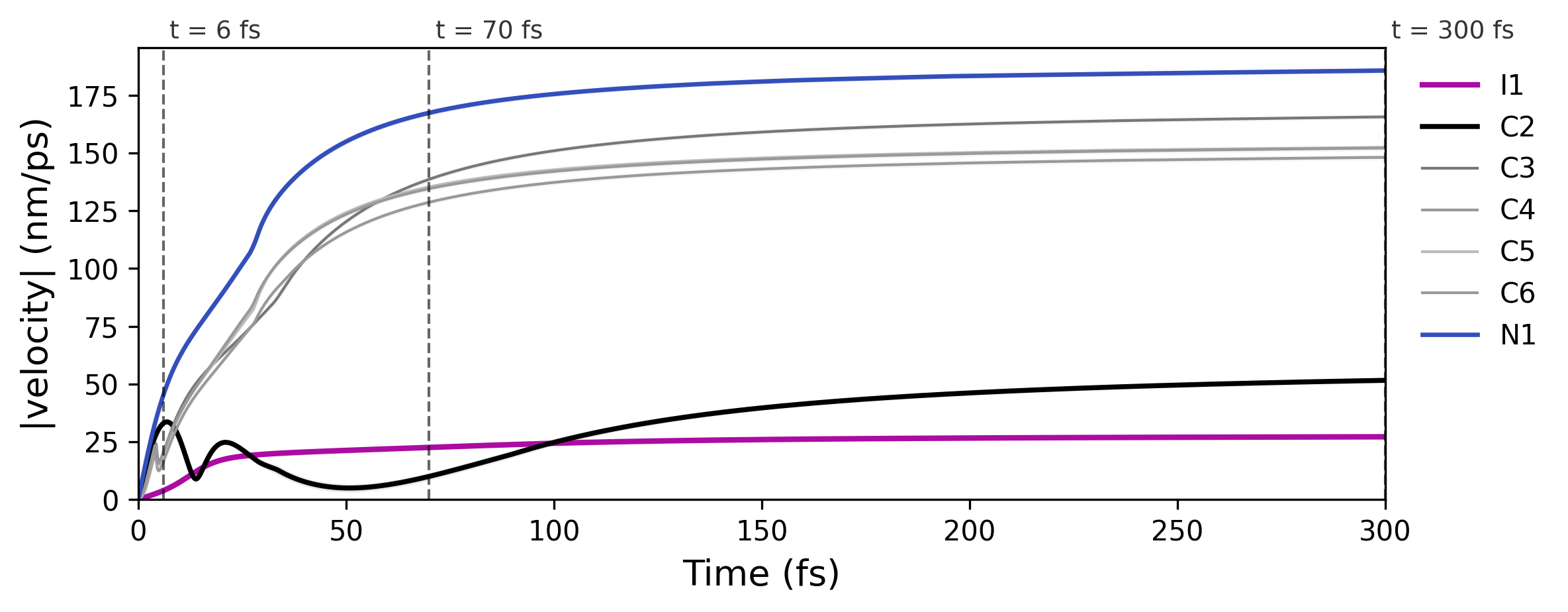}
\caption{Potential and kinetic energy in kJ/mol during the simulated Coulomb explosion of 2-iodopyridine. The dotted lines mark the three timesteps shown in \autoref{fig:2iodpyridine_time_evolution}: at 6~fs the potential energy is still being converted into kinetic energy; at 70~fs the explosion is largely complete and most ions are plateauing; at 300~fs the explosion is fully saturated and all ions have reached their asymptotic momenta.}
\label{fig:2iod-energy}
\end{figure}
~\autoref{fig:2iod-energy} shows the mean fragment speed $\lvert\mathbf{v}i(t)\rvert$ averaged over all 50 trajectories for all heavy atoms. At $t = 300$fs they all have reached their terminal velocities, as evidenced by the vanishing slope of the speed profiles. The momenta extracted at $t = 300$fs therefore represent the asymptotic Coulomb-explosion momenta. 

\subsubsection*{Time evolution of normalised momenta}
\begin{figure*}[h]
    \centering
    \includegraphics[width=\textwidth]{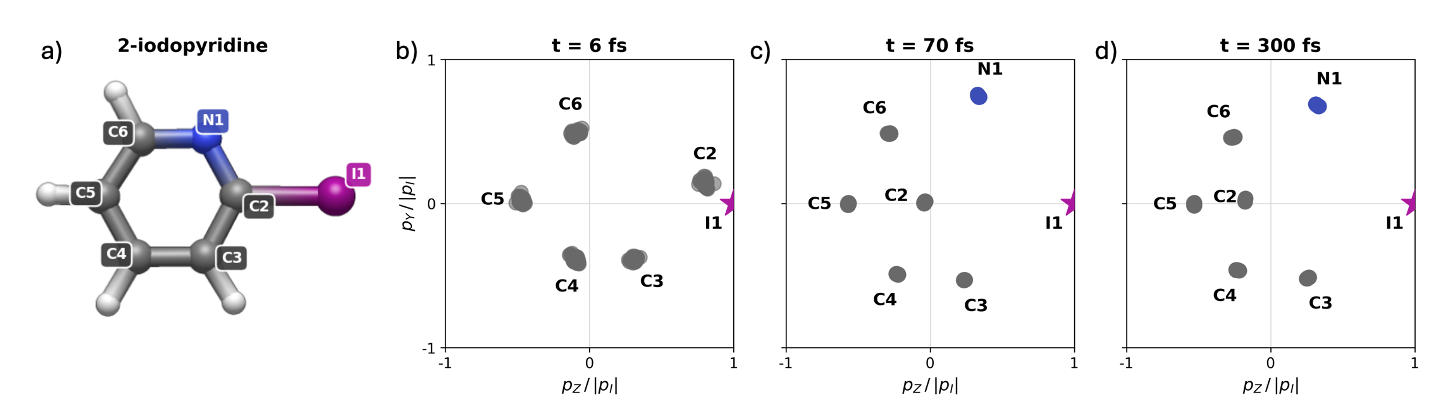}
\caption{\textbf{a)} The molecular structure of 2-iodopyridine. Hydrogen is marked in white, carbon in grey, nitrogen in blue and iodine in violet. \textbf{b)}, \textbf{c)} and \textbf{d)} Normalised momenta of C (grey) and N (blue) fragment ions at 6~fs, 70~fs and 300~fs. Axes are in units of the iodine recoil momentum $|\mathbf{p}_\mathrm{I}|$ at the same timestep; iodine is plotted as a purple star at (1,1).}
\label{fig:2iodpyridine_time_evolution}
\end{figure*}

The time evolution in~\autoref{fig:2iodpyridine_time_evolution} shows the normalised momenta at three representative time steps. Here each trajectory's momenta are divided by $|\mathbf{p}\mathrm{I}|$ at that same time step, so iodine always falls at $(1,0)$ and fragments with momenta exceeding iodine's lie outside the plotted range. At $t = 6$fs all ions are still accelerating; since iodine is the heaviest fragment it has accumulated the least momentum, making $|\mathbf{p}\mathrm{I}|$ small and placing nitrogen outside the plot. At $t = 70$fs all ions except C2, the carbon directly bonded to iodine, are approaching their asymptotic momenta. C2 appears near the origin because its accumulated momentum is anomalously small: unlike every other ring atom, it remained transiently bound to iodine throughout the early explosion. This bond is physically active even at $+1$ charge, because iodine retains its $\sigma$-bonding electron alongside five lone-pair electrons, and carbon retains its C--I $\sigma$ electron having lost only a delocalised ring $\pi$ electron on ionisation — the covalent bond is genuinely still intact at short range. In \moldstruct this is modelled through the C--I Morse potential. Because C2 is eleven times lighter than iodine, ring Coulomb forces initially accelerate it toward the departing iodine faster than iodine itself moves, momentarily compressing the Morse bond. The resulting restoring force reverses C2's motion and drives a transient oscillation visible in the speed profile below 50,fs. As the C--I separation grows, the Coulomb repulsion increasingly dominates over the weakening Morse attraction. This is modelled in \moldstruct with a irreversible bond-breaking flag that fires once the separation irreversibly exceeds $2r_0 \approx 4.2$\AA\. This mirrors the physical reality that at this distance the bonding-electron overlap has collapsed and the bond is truly broken. From this point C2 re-accelerates slowly under weak residual Coulomb forces from the departing fragments. By $t = 300$fs all ions have reached their asymptotic momenta, with iodine carrying the largest.

\section{Applications}
\label{applications_methods}

\subsection{Application I: Peptide structure classification} 

\paragraph{Pulse Parameters}
The simulated X-ray pulse is modelled as a Gaussian in time with a photon energy of \SI{600}{\electronvolt}, enabling K-shell photoionisation followed by Auger decay for carbon, nitrogen, oxygen.
The pulse peak is centred at \SI{20}{\femto\second} with a full width at half-maximum (FWHM) of \SI{30}{\femto\second} (Gaussian $\sigma =
\SI{12.74}{\femto\second}$).
The total number of photons per pulse is $5\times10^{11}$, delivered into a focal spot of \SI{100}{\nano\meter} diameter, yielding a photon fluence of $\mathcal{F} = 6.4\times10^{13}$~photons\,\textmu m$^{-2}$, an energy fluence of $\mathcal{F} = \SI{6.12e5}{\joule\per\centi\meter\squared}$, and a peak intensity of $I_\mathrm{peak} = \SI{1.92e19}{\watt\per\centi\meter\squared}$.
Each trajectory is propagated for \SI{150}{\femto\second}.

\paragraph{Analysis} 
To quantify structural deviation from the $\alpha$-helical reference, we use the Root Mean Square Deviation (RMSD) and lDDT value. RMSD works as follows: given $N$ atoms, RMSD is computed after optimal superposition of the two structures as
\begin{equation}
    \mathrm{RMSD} = \sqrt{\frac{1}{N}\sum_{i=1}^{N} \left|\mathbf{r}_i - \mathbf{r}_i^{\,\mathrm{ref}}\right|^2},
\end{equation}
where $\mathbf{r}_i$ and $\mathbf{r}_i^{\,\mathrm{ref}}$ are the Cartesian coordinates of atom $i$ in the target and reference structures, respectively. A larger RMSD indicates greater overall structural deviation.

The local Distance Difference Test (lDDT)~\citep{lddt} measures structural similarity at the local level, without requiring global superposition. For each atom, all pairwise distances to neighbouring atoms within a threshold radius are considered. A distance is counted as conserved if its deviation from the corresponding reference distance falls below a given tolerance. The lDDT score is the average fraction of conserved distances, computed over four tolerance thresholds (0.5, 1, 2, and 4~\AA). Scores range from 0 to 1, where 1 indicates perfect local conservation of interatomic distances.

Unlike lDDT, which evaluates local distance differences without requiring global superposition, RMSD measures the global displacement of all atoms after rigid-body alignment, making it sensitive to large-scale conformational changes such as the helix-to-extended transition of $\mathrm{Ala_{16}}$. 

We compare the separability of the explosion patterns of an ensemble of structures compared to the $\alpha$-helical confirmation. This ensemble was collected from the unfolding process of $\mathrm{Ala_{16}}$, induced by temperature to the $\alpha$-helical confirmation as well as of the extended confirmation. To quantify cluster separation, the axis in the PCA embedding along which the inter-cluster distance was maximal was determined, and all points were projected onto this axis.

\paragraph{Results}

\begin{figure}[p]
    \centering
    \setlength{\tabcolsep}{0pt} % remove spacing in tabular

    \begin{tabular}[t]{ccc}
        \includegraphics[height=0.95\textheight, keepaspectratio]{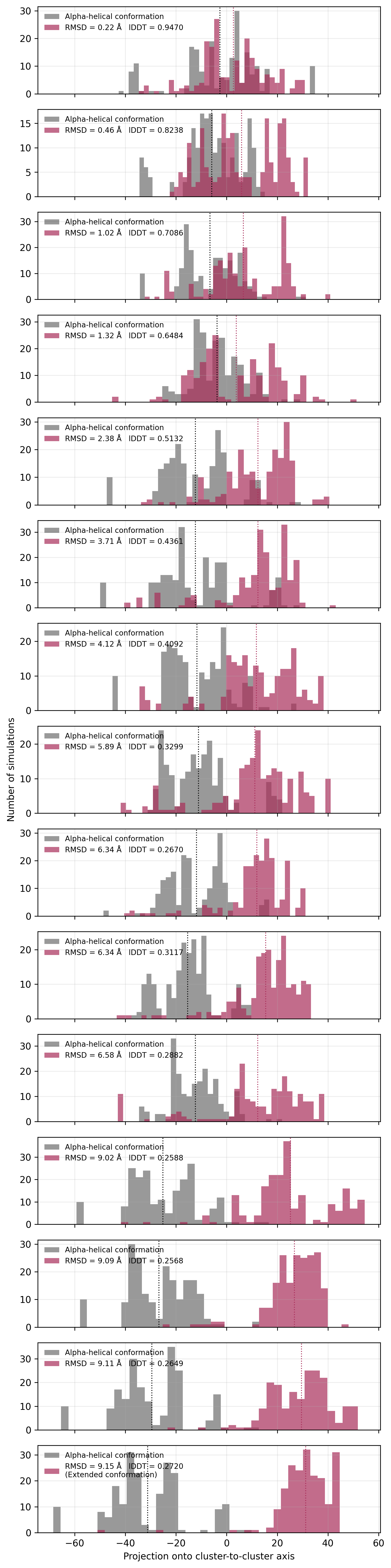} &
        \includegraphics[height=0.95\textheight, keepaspectratio]{images/supplementary/16Ala_target_ref_RMSD.png}  &
        \includegraphics[height=0.95\textheight, keepaspectratio]{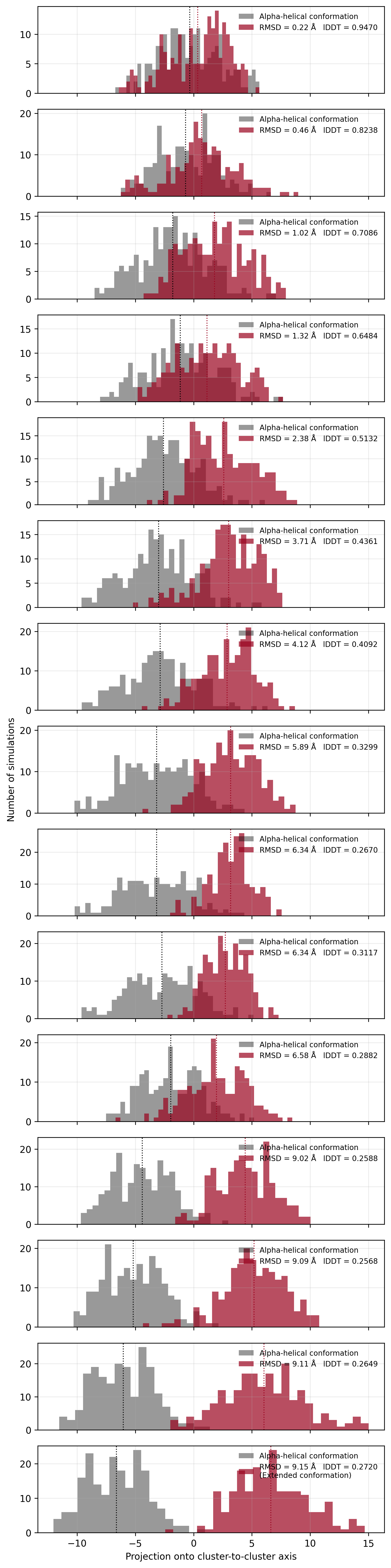}
    \end{tabular}

    \caption{Conformer separability of $\mathrm{Ala_{16}}$ ordered by RMSD from the $\alpha$-helical reference. Left: t-SNE cluster projections of Coulomb-explosion ion-hit maps. Centre: corresponding molecular structures, with the reference coloured in black and the target conformer in blue-to-red. Right: PCA cluster projections of the same data. This is essentially the same data as presented in ~\autoref{fig:tsne_pca_projection_plots_lddt}, however ordered according to RMSD not lDDT.}
    \label{fig:tsne_pca_projection_plots_rmsd}
\end{figure}

~\autoref{fig:tsne_pca_projection_plots_rmsd} presents the ensemble of $\mathrm{Ala_{16}}$ conformers and their separability via Coulomb explosion analysis, using t-SNE and PCA for dimensionality reduction, ordered according to their RMSD from the reference structure. Again, the centre column shows each analysed structure alongside the $\alpha$-helical reference structure (grey). 

The plots show that separability correlates with higher RMSD value, starting from 9.09 ~\AA\ the conformers are separable using both PCA or t-SNE.

\begin{figure}[p]
    \centering
    \setlength{\tabcolsep}{0pt} % remove spacing in tabular

    \begin{tabular}[t]{ccc}
        \includegraphics[height=0.95\textheight, keepaspectratio]{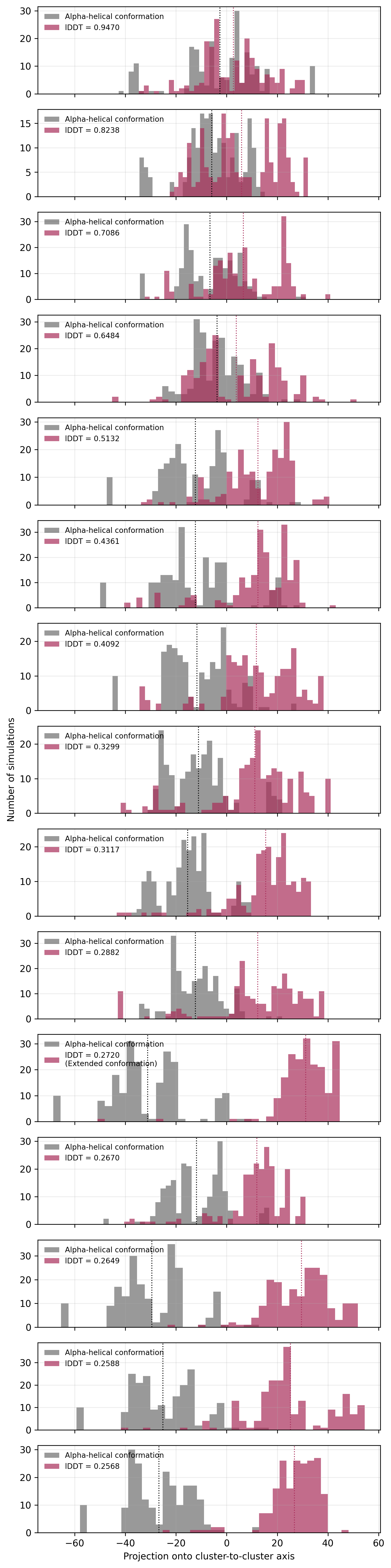} &
        \includegraphics[height=0.95\textheight, keepaspectratio]{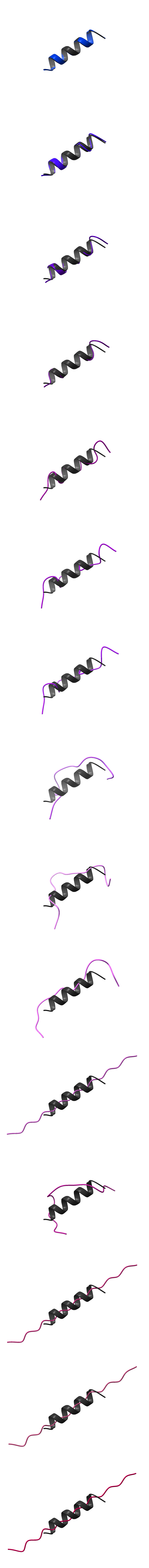}  &
        \includegraphics[height=0.95\textheight, keepaspectratio]{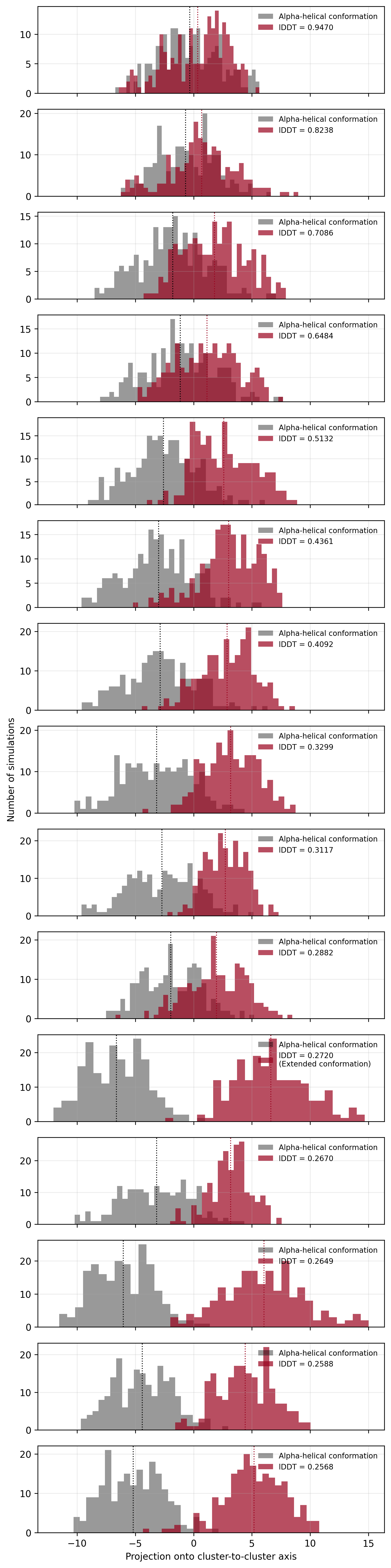}
    \end{tabular}

    \caption{Conformer separability of $\mathrm{Ala_{16}}$ ordered by lDDT from the $\alpha$-helical reference. Left: t-SNE cluster projections of Coulomb-explosion ion-hit maps. Centre: corresponding molecular structures, with the reference coloured in black and the target conformer in blue-to-red. Right: PCA cluster projections of the same data.}
    \label{fig:tsne_pca_projection_plots_lddt}
\end{figure}

~\autoref{fig:tsne_pca_projection_plots_lddt} shows the projection of the separability of the structures from the reference structure ordered with respect to their lDDT value, going from a high value on the top signifying high similarity to a low lDDT value on the bottom. It is essentially the same data as in ~\autoref{fig:tsne_pca_projection_plots_rmsd}, however ordered according to lDDT not RMSD. On the left, the analysis is done using the dimensional reduction algorithm t-SNE and on the right PCA. While the separability does not increase monotonically from high to low lDDT, a clear trend towards greater separability at lower lDDT values is nonetheless apparent, with some outliers.

\subsection{Application II: Ubiquitin structure classification}
\paragraph{Pulse Parameters}
The simulated X-ray pulse is modelled as a Gaussian in time with a photon
energy of \SI{600}{\electronvolt}, above the K-shell ionisation thresholds of
carbon (\SI{284}{\electronvolt}), nitrogen (\SI{410}{\electronvolt}), and
oxygen (\SI{543}{\electronvolt}).
The pulse peak is centred at \SI{30}{\femto\second} with a full width at
half-maximum (FWHM) of \SI{30}{\femto\second} (Gaussian $\sigma =
\SI{12.74}{\femto\second}$), representing a coupled ionisation--explosion
regime in which nuclear motion and charge rearrangement occur on the same
timescale.
The total number of photons per pulse is $1\times10^{11}$, delivered into a
focal spot of \SI{100}{\nano\meter} diameter, yielding a photon fluence of
$\mathcal{F} = 1.27\times10^{13}$~photons\,\textmu m$^{-2}$, an energy
fluence of $\mathcal{F} = \SI{1.22e5}{\joule\per\centi\meter\squared}$, and a
peak intensity of $I_\mathrm{peak} =
\SI{3.83e18}{\watt\per\centi\meter\squared}$.
Each trajectory is propagated for \SI{200}{\femto\second}.

%\nocite{*}
\bibliographystyle{unsrt}
\bibliography{bib}% Produces the bibliography via BibTeX.

\end{document}